\RequirePackage{fixltx2e}
\documentclass[a4paper,fleqn,reqno,english,journal=cmatex,manuscript=article]{revtex4-2}
\usepackage[LGR,T1]{fontenc}
\usepackage[latin9]{inputenc}
\usepackage{babel}
\usepackage{array}
\usepackage{float}
\usepackage{booktabs}
\usepackage{textcomp}
\usepackage{multirow}
\usepackage{amsmath}
\usepackage{amsthm}
\usepackage{amssymb}
\usepackage{graphicx}
\usepackage[unicode=true,pdfusetitle,
 bookmarks=true,bookmarksnumbered=false,bookmarksopen=false,
 breaklinks=false,pdfborder={0 0 1},backref=false,colorlinks=false]
 {hyperref}

\makeatletter

\pdfpageheight\paperheight
\pdfpagewidth\paperwidth

\DeclareRobustCommand{\greektext}{%
  \fontencoding{LGR}\selectfont\def\encodingdefault{LGR}}
\DeclareRobustCommand{\textgreek}[1]{\leavevmode{\greektext #1}}
\ProvideTextCommand{\~}{LGR}[1]{\char126#1}

\providecommand{\tabularnewline}{\\}

\usepackage{lmodern}

\makeatother

\begin{document}
\title{Unveiling the Optoelectronic Potential of Vacancy-Ordered Double Perovskites:
A Computational Deep Dive}
\author{Surajit Adhikari{*}, Ayan Chakravorty, and Priya Johari}
\email{sa731@snu.edu.in, priya.johari@snu.edu.in}

\affiliation{Department of Physics, School of Natural Sciences, Shiv Nadar Institution
of Eminence, Greater Noida, Gautam Buddha Nagar, Uttar Pradesh 201314,
India}
\begin{abstract}
Lead-free perovskite materials have emerged as key players in optoelectronics,
showcasing exceptional optical and electronic properties, alongside
being environmentally friendly and non-toxic elements. Recently, among
studied perovskite materials, vacancy-ordered double perovskites (VODPs)
stand out as a promising alternative. In this study, we captured the
electronic, optical, excitonic, and polaronic properties of a series
of VODPs with the chemical formula Rb$_{2}$BX$_{6}$ (B = Si, Ge,
Sn, Pt; X = Cl, Br, I) using first-principles calculations. Our results
indicate these materials exhibit high stability and notable electronic
and optical properties. The calculated G$_{0}$W$_{0}$ bandgap values
of these perovskites fall within the range of 0.56 to 6.12 eV. Optical
properties indicate strong infra-red to ultraviolet light absorption
across most of the systems. Additionally, an analysis of excitonic
properties reveals low to moderate exciton-binding energies and variable
exciton lifetimes, implying higher quantum yield and conversion efficiency.
Furthermore, utilizing the Feynman polaron model, polaronic parameters
are evaluated, and for the majority of systems, charge-separated polaronic
states are less stable than bound excitons. Finally, an investigation
of Polaronic mobility reveals high polaron mobility for electrons
(3.33$-$85.11 cm$^{2}$V$^{-1}$s$^{-1}$) compared to previously
reported Cs-based VODP materials. Overall, these findings highlight
Rb-based VODPs as promising candidates for future optoelectronic applications.
\end{abstract}
\maketitle

\section{Introduction:}

The development of halide perovskites (HPs) for optoelectronic applications
has made tremendous progress in recent years. Primitively, organic-inorganic
lead-halide perovskites APbX$_{3}$(A = Rb, Cs, CH$_{3}$NH$_{3}$;
X = Cl, Br, and I) drew a lot of attention due to their strong optical
absorption, variable energy gap, long carrier diffusion length, and
excellent power conversion efficiencies\citep{21}. Due to their exceptional
optoelectronic properties, organic-inorganic lead halide perovskites
are highly sought after for a range of applications, including efficient
solar cells, information storage, humidity sensors, LEDs, lasers,
and photoelectric detectors. The power conversion efficiency (PCE)
of these perovskite solar cells has reached up to 24.2\%\citep{92},
surpassing some silicon-based thin-film solar cells. Consequently,
organic-inorganic lead halide perovskites have emerged as the most
extensively researched and cost-effective materials for photovoltaic
energy conversion in recent years.

Despite their outstanding performance, the practical applications
of CH$_{3}$NH$_{3}$-based materials are limited due to the volatility
and moisture sensitivity of CH$_{3}$NH$_{3}$, as well as the toxic
nature of lead (Pb), and their poor long-term stability. This highlights
the need to explore Pb-free hybrid halide perovskites.\citep{49}.
To get rid of lead, a new approach has been developed where two Pb$^{2+}$
cations are replaced by a combination of one monovalent cation (M$^{+}$)
and one trivalent cation (M$^{3+}$). This substitution gives rise
to a new class of perovskites, called double perovskites with the
formula A$_{2}$M(I)M(III)X$_{6}$, where A represents a monovalent
cation, and X represents halides such as Cl, Br, or I. Several Pb-free
double halide perovskites have been explored in past few years, such
as Cs$_{2}$MSbX$_{6}$ (M = Cu, Ag, Na, K, Rb, Cs, and X = Cl, Br)\citep{93}
and Cs$_{2}$YCuX$_{6}$ (X = Cl, Br, I)\citep{94}. These lead-free
halide double perovskite materials have shown to exhibit outstanding
optoelectronic properties such as high optical absorption, tunable
bandgap, broad absorption spectrum, small carrier effective masses,
long charge diffusion lengths and high charge carrier mobility\citep{70,71}.

While these lead-free alternatives show promise, they still require
significant efficiency improvements to match the performance of lead-halide
perovskites. One major challenge of lead-free halide double perovskites
is their typically large and indirect bandgaps, which hinder their
efficiency in comparison to lead-halide perovskites. This limitation
restricts their potential for applications like high-efficiency solar
cells and optoelectronic devices. To address these limitations, vacancy-ordered
double perovskites (VODPs), a class of double perovskites, with the
formula A$_{2}$BX$_{6}$ have emerged as a more efficient alternative.
These complexes are eco-friendly, containing interesting luminescent,
optical, dielectric, and magnetic properties and have applications
in modern semiconductor devices\citep{4}. The VODP materials can
be viewed as a derivative structure of the traditional ABX$_{3}$
by orderly deleting half of the B atoms in the core of the octahedron
in a regular motif\citep{2}. VODPs, with B in a \textquoteleft +4\textquoteright{} oxidation state,
demonstrate enhanced stability in the presence of air and moisture
compared to traditional ABX$_{3}$ perovskites\citep{4,7}. This structural
modification expands the composition space, allowing them for chemical
tuning and property optimization. The transition from traditional
single and double halide perovskites {[}ABX$_{3}$ and A$_{2}$M(I)M(III)X$_{6}${]}
to A$_{2}$BX$_{6}$, thus presents exciting prospects for advancing
optoelectronic materials with broader applications. In contrast, there
is relatively little room for chemical modification in the halide
ABX$_{3}$ structure, and only a few halide perovskites can form;
those have already been thoroughly studied\citep{8}.

Recent research based on the A$_{2}$BX$_{6}$ has opened new possibilities
for a stabilized and eco-friendly solar cell material\citep{9}. Compounds
such as Rb$_{2}$PdI$_{6}$, Rb$_{2}$PdBr$_{6}$, and Cs$_{2}$PtI$_{6}$
exhibit ideal bandgaps, high dielectric constants, and strong absorption
coefficients, making them promising for solar cell applications due
to their robust nature and environmental benefits\citep{61}. Zhao
et al. used DFT-based first-principles calculations to examine the
electronic and optical properties of inorganic A$_{2}$PtI$_{6}$
(A = Rb or Cs) compounds\citep{11}, which having bandgaps of 1.15
eV and 1.29 eV, suitable for solar cells and optoelectronic devices.
Jiang et al. determined the direct bandgaps and computed the mechanical,
thermal, electrical, and optical properties of Cs$_{2}$SnX$_{6}$
(X = Cl, Br, and I)\citep{14}. Several materials, including Cs$_{2}$TeI$_{6}$,
Cs$_{2}$SnI$_{6}$, and Cs$_{2}$TiI$_{6}$, with suitable bandgaps
are known in theory and experiment\citep{16,17,2}. In the context
of DFT and hybrid functional (HSE06), the stability and electronic
structure of A$_{2}$BX$_{6}$ (A = K, Rb, Cs; B = Si, Ge, Sn, Pb,
Ni, Pd, Pt, Se, Te; and X = Cl, Br, I) have been studied \citep{20}.

The synthesis of thin films and single crystals of Cs$_{2}$TeI$_{6}$
has been done having an optical bandgap of 1.5 eV and Cs$_{2}$PdBr$_{6}$
showing a bandgap of 1.69 eV and strong long-term stability\citep{19,35}.
Additionally, microcrystals of Rb$_{2}$HfCl$_{6}$ and various Rb$_{2}$TeX$_{6}$(X=Cl,
Br, I) compounds have been experimentally synthesized\citep{4,32}.
Recent advancements include the synthesis of Rb$_{2}$SnCl$_{6}$,
Rb$_{2}$SnBr$_{6}$, K$_{2}$SnBr$_{6}$, and Rb$_{2}$SnI$_{6}$\citep{38}.
Recently, the potential for using Cs$_{2}$SnI$_{6}$ in solar technology
has also been investigated\citep{2}. The optoelectronic properties
of various VODPs are significantly influenced by the B-site cation,
with the choice of element affecting their structural and tunable
optoelectronic characteristics due to the {[}BX$_{6}${]}$^{-}$ octahedra.
However, Sn$^{2+}$ and Ge$^{2+}$ cations are prone to oxidation
into Sn$^{4+}$ and Ge$^{4+}$, which can lead to instability in perovskite
solar cells (PSCs) based on these materials. Despite this challenge,
Cs$_{2}$SnI$_{6}$, with its unique VODP structure, shows potential
as a light-absorbing layer for PSCs\citep{95}. This underscores the
necessity of exploring Sn- and Ge-based VODPs further to overcome
stability issues and optimize their performance in optoelectronic
applications.. Another fact is that crystalline Si solar cells are
widely used in industry but are more expensive to produce, which has
prompted researchers to investigate the theoretical features of Si-based
DPs. Cs$_{2}$SiI$_{6}$has already been studied computationally and
it possesses a direct bandgap of 2.72 eV. Till now, according to our
report, Rb$_{2}$SiBr$_{6}$ and Rb$_{2}$GeBr$_{6}$ have yet to
be studied experimentally and theoretically.

In this contribution, we employed first-principles density functional
theory (DFT) and many-body perturbation theory (MBPT) based simulations
to extensively explore the phase stability and optoelectronic properties
of lead (Pb)-free Rb$_{2}$BX$_{6}$ (B = Si, Ge, Sn, Pt; X = Cl,
Br, I) VODPs. All compounds maintain a stable, standard cubic double
perovskite structure with alternating B-site vacancies, showing promising
mechanical stability and flexibility. Electronic properties reveal
that these materials have a direct bandgap except Rb$_{2}$PtBr$_{6}$
and Rb$_{2}$PtI$_{6}$. Then, utilizing excited-state methods, we
computed the optical characteristics using the GW-BSE methodology
based on the many-body perturbation theory (MBPT). The Bethe-Salpeter
equation (BSE)\citep{45,26}, on top of a single-shot GW(G$_{0}$W$_{0}$)@PBE\citep{46},
has been solved to ascertain the electronic contribution to the dielectric
function. Further, the ionic contribution to the dielectric function
is obtained using density functional perturbation theory (DFPT). Aditionally,
our results indicate low carrier effective mass, high mobility, varied
exciton lifetimes, and low to moderate exciton binding energies. Finally,
we determined several polaronic properties, suggesting that most of
the materials have high polaron mobility facilitating efficient charge
transport in optoelectronic devices. Overall, this study offers a
thorough analysis of a number of Rb$_{2}$BX$_{6}$ compounds and
creates a pathway for the intriguing lead-free perovskite possibilities,
which have significant promise for next-generation optoelectronic
and photovoltaic devices.

\section{Computational Details :}

In our present study, first-principles density-functional theory (DFT)
and many-body perturbation theory (MBPT) based computations were carried
out using the Vienna ab initio simulation package (VASP)\citep{22,23}.
The projected augmented wave (PAW) approach was employed in all calculations
to accurately model the interactions between the valence electrons
and the atomic core\citep{24}. The PAW pseudopotentials with valence
shell electrons of each atom were being considered as follows: 9 for
Rb -4s$^{2}$4p$^{6}$5s$^{1}$, 4 for Si -3s$^{2}$3p$^{2}$, 14
for Ge - 3d$^{10}$4s$^{2}$4p$^{2}$, 14 for Sn - 4d$^{10}$5s$^{2}$5p$^{2}$,
10 for Pt - 5d$^{9}$6s$^{1}$, 7 for Cl - 3s$^{2}$3p$^{5}$, 7 for
Br - 4s$^{2}$4p$^{5}$, and 7 for I - 5s$^{2}$5p$^{5}$. The exchange-correlation
(xc) functional of Perdew, Burke, and Ernzerhof (PBE), which is based
on the generalized gradient approximation (GGA), was used to account
for the electron-electron interactions\citep{25}. Utilizing the conjugate
gradient approach, the total energy was minimized to generate the
optimal structural parameters for each of the VODP system. The calculations
for all the systems were performed using an optimized plane-wave cutoff
energy of 400 eV. The convergence conditions for self-consistent-field
iteration and geometry relaxation in all calculations were set at
$10^{-6}$ eV for energy and 0.01 eV/$\textrm{\AA}$ for Hellmann-Feynman
forces on each atom. PBE xc functional with \textgreek{G}-centered
\textbf{k}-point of $4\times4\times4$ grid was used for structural
optimization. The PBE functional, however, are known to underestimate
energy bandgaps and optical characteristics due to the self-interaction
error. To obtain more accurate electronic structures, hybrid Heyd\textminus Scuseria\textminus Ernzerhof
(HSE06)\citep{26} and many-body perturbation theory (MBPT) based
G$_{0}$W$_{0}$@PBE\citep{29,30} calculations were performed for
all the systems. Note that the band structures were calculated using
the PBE xc functionals, taking into account the spin-orbit coupling
effect. In addition, the density of state (DOS) computations were
performed using the hybrid HSE06 functional with \textgreek{G}-centered
$6\times6\times6$ \textbf{k}-point grid. To ascertain accurate optical
properties, we solved the MBPT-based Bethe-Salpeter equation (BSE)\citep{27,28}
on top of single-shot GW(G$_{0}$W$_{0}$)@PBE\citep{29,30} computations.
In the GW-BSE calculations, we utilized \textgreek{G}-centered $4\times4\times4$
\textbf{k}-grid, achieving complete convergence with an impressive
number of NBANDS, specifically 540. The electron-hole kernel for BSE
calculations was constructed using 6 occupied and 6 unoccupied states.
For each system, the ionic contribution to the dielectric function
was computed using density functional perturbation theory (DFPT)\citep{31}
with a $4\times4\times4$ \textbf{k-}point grid.

\section{Results and Discussions:}

In this study, we performed a thorough and systematic investigation
of vacancy-ordered double perovskites Rb$_{2}$BX$_{6}$ (B = Si,
Ge, Sn, Pt; X = Cl, Br, I) to evaluate their potential optoelectronic
properties. The following sections delve into a detailed analysis
of the stability, along with the structural, electronic, optical,
excitonic, and polaronic properties of these Rb$_{2}$BX$_{6}$ compounds,
offering fundamental insights to inform and guide future experimental
research.

\subsection{Structural Properties:}

\subsubsection{Crystal Structure:}

The Rb$_{2}$BX$_{6}$ double perovskites, having elements Rb, Si,
Ge, Sn, Pt, Cl, Br, and I, showcase an orderly face-centered cubic
crystal structure designated by the space group Fm$\bar{3}$m (225).
This configuration is similar to insufficient B-site ABX$_{3}$ perovskite
materials that have discrete {[}BX$_{6}${]} clusters. In this configuration,
12 halogen ions surround the A-site cations (Rb), while 6 halogen
ions coordinate with the B-site atoms.

In the crystal structure, Rb atoms find their place at specific 8c
Wyckoff positions with coordinates (0.25, 0.25, 0.25). Meanwhile,
B atoms occupy 4a Wyckoff positions at (0, 0, 0), while X atoms are
located at 24e Wyckoff positions with coordinates (x, 0, 0), where
x hovers around 0.20. In this arrangement, Rb atoms reside between
{[}BX$_{6}${]} octahedra, surrounded by 12 halogen atoms, while B
atoms take position at the corners and face-centered spots of {[}BX$_{6}${]}
octahedra.

\begin{figure}[H]
\begin{centering}
\includegraphics[width=1\textwidth,height=1\textheight,keepaspectratio]{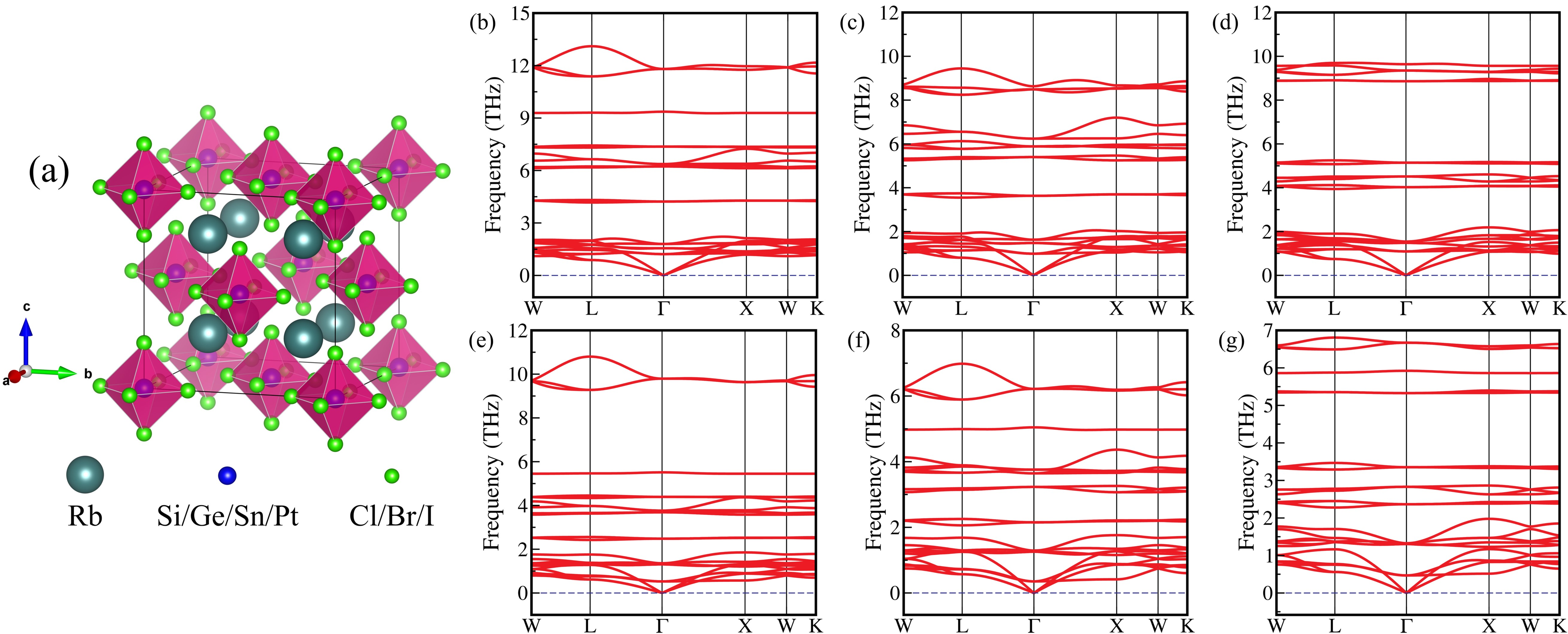}
\par\end{centering}
\caption{\label{fig:1}(a) Crystal structure of Rb$_{2}$BX$_{6}$ (B = Si,
Ge, Sn, Pt; X = Cl, Br, I) VODPs, and phonon dispersion curves of
(b) Rb$_{2}$SiCl$_{6}$, (c) Rb$_{2}$GeCl$_{6}$, (d) Rb$_{2}$PtCl$_{6}$,
(e) Rb$_{2}$SiBr$_{6}$, (f) Rb$_{2}$GeBr$_{6}$, and (g) Rb$_{2}$PtBr$_{6}$
VODP, respectively, calculated with the DFPT method.}
\end{figure}

Table \ref{tab:1} provides an overview of the optimized lattice parameters
of the analyzed VODPs. We find that our computed lattice parameters
match well with the earlier theoretical and experimental data\citep{59,60,61,62,63}.
Based on our calculations, we observe a consistent pattern of the
lattice constant increasing gradually when transitioning from lower
to higher atomic radius of the halogen atoms. We also notice a comparable
trend in bond lengths, where the distances for both Rb-X and B-X bonds
consistently increase as we progress from chlorine (Cl) to iodine
(I).

\begin{table}[H]
{\small{}\caption{\label{tab:1}Calculated lattice parameters, bond lengths and formation
energies ($E_{f}$) of Rb$_{2}$BX$_{6}$ (B = Si, Ge, Sn, Pt; X =
Cl, Br, I) VODPs.}
}{\small\par}
\begin{centering}
{\footnotesize{}}%
\begin{tabular}{cccccccc}
\hline 
\multirow{2}{*}{{\footnotesize{}Configurations}} & \multicolumn{2}{c}{{\footnotesize{}Lattice parameter (a = b = c), $\mathring{A}$}} &  & \multicolumn{2}{c}{{\footnotesize{}Bond length, $\mathring{A}$}} &  & {\footnotesize{}Formation energy ($E_{f}$)}\tabularnewline
\cline{2-3} \cline{3-3} \cline{5-6} \cline{6-6} 
 & {\footnotesize{}In our study} & {\footnotesize{}Previous report} &  & {\footnotesize{}Rb$-$X} & {\footnotesize{}B$-$X} &  & {\footnotesize{}(eV/atom)}\tabularnewline
\hline 
{\footnotesize{}Rb$_{2}$SiCl$_{6}$} & {\footnotesize{}10.12} &  &  & {\footnotesize{}3.59} & {\footnotesize{}2.24} &  & {\footnotesize{}-1.592}\tabularnewline
{\footnotesize{}Rb$_{2}$SiBr$_{6}$} & {\footnotesize{}10.71} &  &  & {\footnotesize{}3.79} & {\footnotesize{}2.43} &  & {\footnotesize{}-1.111}\tabularnewline
{\footnotesize{}Rb$_{2}$SiI$_{6}$} & {\footnotesize{}11.67} &  &  & {\footnotesize{}4.13} & {\footnotesize{}2.70} &  & {\footnotesize{}-0.926}\tabularnewline
{\footnotesize{}Rb$_{2}$GeCl$_{6}$} & {\footnotesize{}10.22} &  &  & {\footnotesize{}3.62} & {\footnotesize{}2.33} &  & {\footnotesize{}-1.403}\tabularnewline
{\footnotesize{}Rb$_{2}$GeBr$_{6}$} & {\footnotesize{}10.77} &  &  & {\footnotesize{}3.81} & {\footnotesize{}2.53} &  & {\footnotesize{}-0.983}\tabularnewline
{\footnotesize{}Rb$_{2}$GeI$_{6}$} & {\footnotesize{}11.64} &  &  & {\footnotesize{}4.12} & {\footnotesize{}2.78} &  & {\footnotesize{}-0.848}\tabularnewline
{\footnotesize{}Rb$_{2}$SnCl$_{6}$} & {\footnotesize{}10.43} & {\footnotesize{}10.12\citep{60}(b)} &  & {\footnotesize{}3.69} & {\footnotesize{}2.47} &  & {\footnotesize{}-1.501}\tabularnewline
{\footnotesize{}Rb$_{2}$SnBr$_{6}$} & {\footnotesize{}10.99} & {\footnotesize{}10.58\citep{62}(a), 11.01\citep{61}(b)} &  & {\footnotesize{}3.89} & {\footnotesize{}2.65} &  & {\footnotesize{}-1.096}\tabularnewline
{\footnotesize{}Rb$_{2}$SnI$_{6}$} & {\footnotesize{}11.81} & {\footnotesize{}11.62\citep{63}(a), 11.87\citep{61}(b)} &  & {\footnotesize{}4.18} & {\footnotesize{}2.91} &  & {\footnotesize{}-0.962}\tabularnewline
{\footnotesize{}Rb$_{2}$PtCl$_{6}$} & {\footnotesize{}10.21} & {\footnotesize{}9.88\citep{84} (a), 10.23\citep{83} (b)} &  & {\footnotesize{}3.61} & {\footnotesize{}2.35} &  & {\footnotesize{}-1.257}\tabularnewline
{\footnotesize{}Rb$_{2}$PtBr$_{6}$} & {\footnotesize{}10.72} & {\footnotesize{}10.72\citep{83} (b)} &  & {\footnotesize{}3.79} & {\footnotesize{}2.50} &  & {\footnotesize{}-0.911}\tabularnewline
{\footnotesize{}Rb$_{2}$PtI$_{6}$} & {\footnotesize{}11.44} & {\footnotesize{}11.22\citep{59}(a), 11.54\citep{85} (b)} &  & {\footnotesize{}4.05} & {\footnotesize{}2.71} &  & {\footnotesize{}-0.881}\tabularnewline
\hline 
\end{tabular}{\footnotesize\par}
\par\end{centering}
\begin{raggedright}
{\small{}a = Experimental reports}{\small\par}
\par\end{raggedright}
\raggedright{}{\small{}b = Theoretical reports}{\small\par}
\end{table}

Furthermore, to predict whether these crystals are stable perovskite
structures, we examined crystallographic stability. We calculated
the Goldschmidt tolerance factor ($t$), octahedral factor ($\mu$),
and new tolerance factor ($\tau$) (For details, see Section I of
the Supplemental Material). The calculated results indicate that these
perovskites remain stable in cubic structures.

\subsubsection{Dynamical Stability:}

We employed the DFPT\citep{64} approach to thoroughly assess the
dynamical stability of the Rb$_{2}$BX$_{6}$ (B = Si, Ge, Sn, Pt;
X = Cl, Br, I) VODPs. The dynamical stability of a material is a key
factor in determining its overall stability as it is closely related
to the characteristics of its phonon modes. In the case of Rb$_{2}$BX$_{6}$,
there are 27 phonon modes linked to its 9 atoms. Among them, 3 are
low-frequency vibrations denoted by acoustic mode, while the others
are high-frequency vibrations denoted by optical mode. Given that
there are no imaginary frequencies for Rb$_{2}$SiCl$_{6}$, Rb$_{2}$GeCl$_{6}$,
Rb$_{2}$PtCl$_{6}$, Rb$_{2}$SiBr$_{6}$, Rb$_{2}$GeBr$_{6}$,
and Rb$_{2}$PtBr$_{6}$, indicating these materials appear to be
dynamically stable at T=0 K. Rest of the configurations are not dynamically
stable at T=0 K. Since there are multiple factors that affect stability,
it is more promising to investigate them altogether. Consequently,
we extended our investigation to include the examination of mechanical
and thermodynamic stability for all the compounds, as detailed in
the following sections of this research paper.

\subsubsection{Mechanical Stability and Elastic Properties:}

Stability is one of the crucial aspects to consider when choosing
a material for its future research applications. Therefore, to acquire
a more comprehensive understanding, the mechanical stability and elastic
properties of Rb$_{2}$BX$_{6}$ compounds are also investigated alongside
the crystallographic and dynamical stability. The mechanical stability
and anisotropic behavior of the materials can be described by calculating
the elastic constants. We have calculated the second-order elastic
constants of 12 VODPs with help of the energy-strain approach\citep{51}.
Three distinct elastic constants - $C_{11}$, $C_{12}$, and $C_{44}$
for cubic symmetry are adequate to account for the mechanical stability
and anisotropic characteristics of the crystal. We can determine the
bulk ($B$), shear ($G$), and Young's ($Y$) moduli, Poisson's Ratio
($\nu$), and Zener Anisotropic Factor ($A$) of materials from these
elastic constants (for details, see Section II of the Supplemental
Material).

The computed values of $C_{11}$, $C_{12}$, and $C_{44}$ confirm
that the elastic constants satisfy the mechanical stability criteria,
indicating that these VODPs are mechanically stable (See Table S2).
With the use of the VASPKIT package\citep{52}, the elastic coefficients
($C_{ij}$) are computed using the GGA-PBE xc functional, and the
related bulk modulus ($B$), shear modulus ($G$), Young's modulus
($Y$), and Poisson's ratio ($\nu$) are established using the Voigt-Reuss-Hill
technique\citep{53,54}.

The macroscopic capacity of an object to withstand compression under
external pressure is described by the bulk modulus ($B$). At the
same time, shear modulus ($G$) and Young's modulus ($Y$) describe
the shear resistance and stiffness of the material, respectively.
In our case, Rb$_{2}$GeCl$_{6}$ has the largest bulk and shear moduli
making it the most resistant to compression and shear. Our results
show that shear deformation decreases as the X component changes from
Cl to I. Similarly, flexibility increases with lower Young's modulus,
with iodine-containing materials being more flexible than their Cl
and Br counterparts. Next, Poisson ($\nu$) and Pugh ratios ($B/G$)
are calculated in order to understand the brittle and ductile characteristics
of the materials under consideration. If $\nu>0.26$ and $B/G>1.75$,
it is recognized that the material is ductile; otherwise, it is brittle\citep{55,56,57}.
From our results, we can conclude that except Rb$_{2}$SiCl$_{6}$,
Rb$_{2}$GeCl$_{6}$, and Rb$_{2}$PtCl$_{6}$, all the other perovskite
compounds are ductile and the mentioned three are brittle. Perovskite's
ductile quality makes it possible to employ it to create flexible
optoelectronic devices. So, we can predict that, except for those
three materials, all the other materials can be used for more advanced
flexible optoelectronic devices. In addition to the directional characteristics
of these materials, we have calculated the Zener anisotropic factor
($A$)\citep{58}. The value of $A$ is equal to 1 for isotropic materials,
and any variation from unity in the value of $A$ indicates that the
material has a high degree of anisotropy. Our results reveal that
all the considered configurations are anisotropic in nature. After
analyzing these 12 different perovskite materials, we find they have
different physical and chemical characteristics in various orientations.
To make the most of perovskites for things like solar cells and other
optoelectronic devices, it's crucial to figure out how to utilize
them in real-world scenarios best.

\subsubsection{Thermodynamical Stability:}

In assessing the thermodynamic stability of the compounds, the formation
energy ($E_{f}$) is calculated using the following formula:
\begin{center}
\begin{equation}
E_{f}=\frac{E_{Rb_{l}B_{m}X_{n}}-lE_{Rb}-mE_{B}-nE_{X}}{(l+m+n)}
\end{equation}
\par\end{center}

where $E_{Rb_{l}B_{m}X_{n}}$ represents the total energy of investigated
compounds and $E_{Rb}$, $E_{B}$, and $E_{X}$ are the energies of
individual Rb, B-site (Si, Ge, Sn, Pt) and halogen (Cl, Br, I) atoms
in the crystal structure, respectively. Here, $l$, $m$, $n$ refers
to the number of atoms ($l$=2, $m$=1, $n$=6). Exothermic reactions
release heat, which is often seen as advantageous for practical synthesis.
A negative value of $E_{f}$ indicates an exothermic reaction, indicating
the possibility of the compound's synthesis. The $E_{f}$ values of
these VODPs are depicted in Table \ref{tab:1}. Hence, it can be concluded
that the synthesis of these compounds is indeed feasible.

\subsection{Electronic Properties:}

Once we confirmed the stability of the chosen materials, we moved
on to the next step to analyze the electronic structure of Rb$_{2}$BX$_{6}$
VODPs, which is crucial for building optoelectronic devices. Initially,
GGA-PBE xc functional\citep{25} is used for all computations. We
further employed the hybrid HSE06 xc functional\citep{26} and many-body
perturbation theory (G$_{0}$W$_{0}$) to calculate the energy bandgaps
since PBE functional is well known to underestimate the bandgaps due
to the self-interaction error. The spin-orbit coupling (SOC) effects
are also taken into account while calculating the electronic properties
using the PBE functional. The bandgaps of these 12 combinations are
listed in Table \ref{tab:2}. Estimated bandgap values for different
substances in the Fm$\bar{3}$m phase range from 0.56 eV to 6.12 eV.
Four different elements were taken into account in the current study
at the B site. Though all elements have a ``+4'' oxidation state,
their valence electron orbitals are arranged differently. This results
in variations in their electronic structures, including differences
in the arrangement of energy levels and the types of bandgaps.

\begin{figure}[H]
\begin{centering}
\includegraphics[width=1\textwidth,height=1\textheight,keepaspectratio]{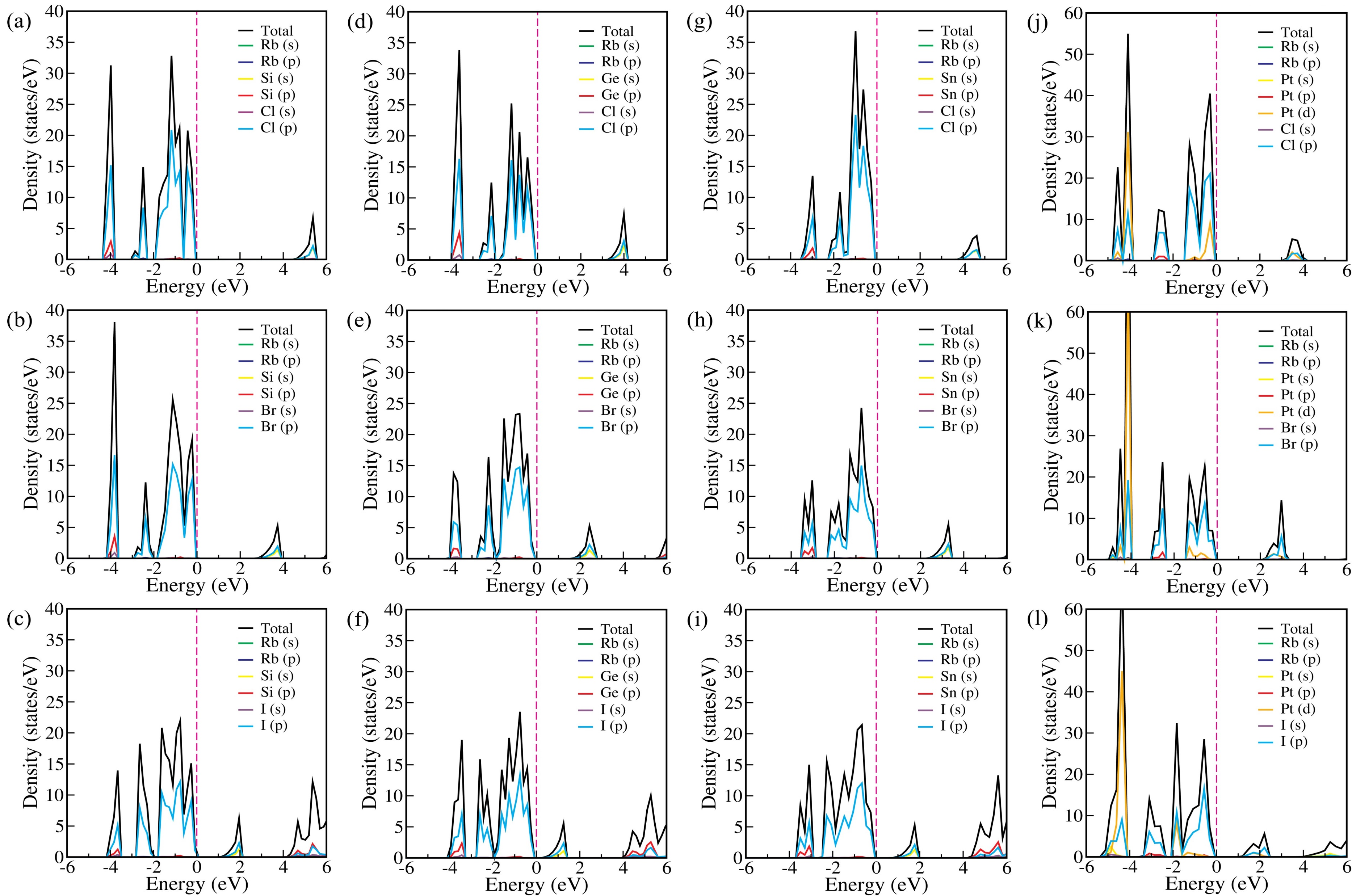}
\par\end{centering}
\caption{\label{fig:2}Calculated electronic total density of states (TDOS)
and partial density of states (PDOS) of (a-c) Rb$_{2}$SiX$_{6}$,
(d-f) Rb$_{2}$GeX$_{6}$, (g-i) Rb$_{2}$SnX$_{6}$, and (j-l) Rb$_{2}$PtX$_{6}$
VODPs using the HSE06 xc functional, where X = Cl, Br, and I, respectively.
The Fermi level is set to be zero and marked by the dashed line.}
\end{figure}

\begin{figure}[H]
\begin{centering}
\includegraphics[width=1\textwidth,height=1\textheight,keepaspectratio]{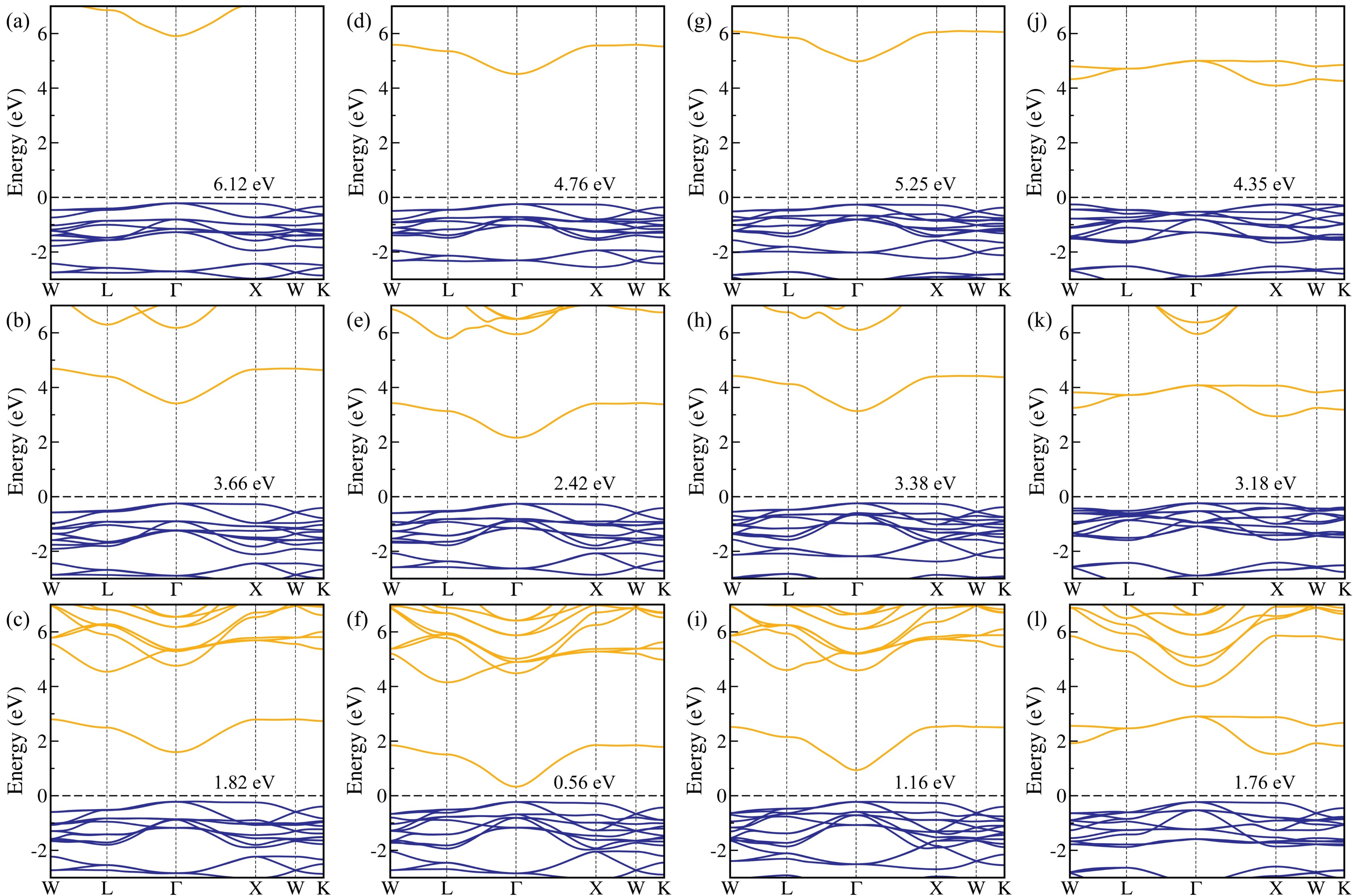}
\par\end{centering}
\caption{\label{fig:3}Calculated electronic band structures of (a-c) Rb$_{2}$SiX$_{6}$,
(d-f) Rb$_{2}$GeX$_{6}$, (g-i) Rb$_{2}$SnX$_{6}$, and (j-l) Rb$_{2}$PtX$_{6}$
VODPs using the G$_{0}$W$_{0}$@PBE method, where X = Cl, Br, and
I, respectively. The Fermi level is set to be zero and marked by the
dashed line.}
\end{figure}

Calculating bandgaps using PBE and PBE-SOC reveals a slight reduction
in the bandgap when SOC is included. The reason for the slight reduction
in the bandgap when SOC is included in calculations is the spin mixing
induced by SOC, which modifies the electronic states near the band
edges, narrowing the bandgap. It changes slightly (< 0.1 eV) when
the B-site cation is Si, Ge, and Sn, and when B is Pt, it changes
up to 0.26 eV. The calculated bandgap values utilizing PBE and HSE06
functionals are consistent with previously reported theoretical values.
In the case of Rb$_{2}$SnI$_{6}$, the energy bandgap value also
agrees with previously reported experimental data. Figure \ref{fig:3}
shows the electronic band structures of Rb$_{2}$BX$_{6}$, computed
using the G$_{0}$W$_{0}$@PBE technique. Except for Rb$_{2}$PtI$_{6}$
and Rb$_{2}$PtBr$_{6}$, all of the compounds have a direct bandgap.
Across all cases, there's a consistent trend: the bandgap values follow
the order I < Br < Cl. This trend can be explained with the help of
Pauling electronegativity\citep{69}. According to Pauling, the electronegativity
values for halogen atoms are in the order of I (2.66) < Br (2.96)
< Cl (3.16). A higher electronegativity of the halogen atom results
in stronger bond interactions between the halogen and B-site atoms,
leading to an upward shift in the conduction band, and consequently,
larger bandgaps.

\begin{table}[H]
\caption{\label{tab:2}Computed bandgap ($E_{g}$) of Rb$_{2}$BX$_{6}$ (B
= Si, Ge, Sn, Pt; X = Cl, Br, I) VODPs using the PBE/PBE-SOC, HSE06
xc-functionals, and\textbf{ }G$_{0}$W$_{0}$@PBE method.}

\begin{centering}
{\scriptsize{}}%
\begin{tabular}{ccccccc}
\hline 
\multirow{2}{*}{{\scriptsize{}Configurations}} & \multicolumn{3}{c}{{\scriptsize{}$E_{g}$ (eV)}} & \multirow{2}{*}{{\scriptsize{}Nature}} & \multicolumn{2}{c}{{\scriptsize{}Previous reports of $E_{g}$ (eV)}}\tabularnewline
\cline{2-4} \cline{3-4} \cline{4-4} \cline{6-7} \cline{7-7} 
 & {\scriptsize{}PBE/PBE-SOC} & {\scriptsize{}HSE06} & {\scriptsize{}G$_{0}$W$_{0}$@PBE} &  & {\scriptsize{}Theoretical} & {\scriptsize{}Experimental}\tabularnewline
\hline 
{\scriptsize{}Rb$_{2}$SiCl$_{6}$} & {\scriptsize{}3.21/3.18} & {\scriptsize{}4.58} & {\scriptsize{}6.12} & {\scriptsize{}D} & {\scriptsize{}4.57\citep{87}(b)} & \tabularnewline
{\scriptsize{}Rb$_{2}$SiBr$_{6}$} & {\scriptsize{}1.64/1.55} & {\scriptsize{}2.74} & {\scriptsize{}3.66} & {\scriptsize{}D} & {\scriptsize{}2.72\citep{87}(b)} & \tabularnewline
{\scriptsize{}Rb$_{2}$SiI$_{6}$} & {\scriptsize{}0.28/0.21} & {\scriptsize{}1.08} & {\scriptsize{}1.82} & {\scriptsize{}D} & {\scriptsize{}0.96\citep{87}(b)} & \tabularnewline
{\scriptsize{}Rb$_{2}$GeCl$_{6}$} & {\scriptsize{}2.04/2.02} & {\scriptsize{}3.37} & {\scriptsize{}4.76} & {\scriptsize{}D} &  & \tabularnewline
{\scriptsize{}Rb$_{2}$GeBr$_{6}$} & {\scriptsize{}0.72/0.66} & {\scriptsize{}1.76} & {\scriptsize{}2.42} & {\scriptsize{}D} &  & \tabularnewline
{\scriptsize{}Rb$_{2}$GeI$_{6}$} & {\scriptsize{}0.01/$-$} & {\scriptsize{}0.33} & {\scriptsize{}0.56} & {\scriptsize{}D} & {\scriptsize{}1.24\citep{88}(c)} & \tabularnewline
{\scriptsize{}Rb$_{2}$SnCl$_{6}$} & {\scriptsize{}2.54/2.52} & {\scriptsize{}3.87} & {\scriptsize{}5.25} & {\scriptsize{}D} & {\scriptsize{}2.40\citep{90}(a)} & {\scriptsize{}3.85\citep{89}}\tabularnewline
{\scriptsize{}Rb$_{2}$SnBr$_{6}$} & {\scriptsize{}1.30/1.22} & {\scriptsize{}2.37} & {\scriptsize{}3.38} & {\scriptsize{}D} & {\scriptsize{}1.30 \citep{61}(a), 2.17\citep{61} (b)} & \tabularnewline
{\scriptsize{}Rb$_{2}$SnI$_{6}$} & {\scriptsize{}0.03/0.01} & {\scriptsize{}0.80} & {\scriptsize{}1.16} & {\scriptsize{}D} & {\scriptsize{}0.12\citep{61} (a), 1.10 \citep{61} (b)} & {\scriptsize{}1.32\citep{65}}\tabularnewline
{\scriptsize{}Rb$_{2}$PtCl$_{6}$} & {\scriptsize{}1.97/1.71} & {\scriptsize{}3.30} & {\scriptsize{}4.35} & {\scriptsize{}D} & {\scriptsize{}3.39\citep{91}(c)} & \tabularnewline
{\scriptsize{}Rb$_{2}$PtBr$_{6}$} & {\scriptsize{}1.28/1.17} & {\scriptsize{}2.26} & {\scriptsize{}3.18} & {\scriptsize{}I} & {\scriptsize{}1.30\citep{20} (b)} & \tabularnewline
{\scriptsize{}Rb$_{2}$PtI$_{6}$} & {\scriptsize{}0.46/0.34} & {\scriptsize{}1.19} & {\scriptsize{}1.76} & {\scriptsize{}I} & {\scriptsize{}1.15 \citep{11}(b)} & \tabularnewline
\hline 
\end{tabular}{\scriptsize\par}
\par\end{centering}
\begin{raggedright}
{\footnotesize{}a = PBE}{\footnotesize\par}
\par\end{raggedright}
\begin{raggedright}
{\footnotesize{}b = HSE06}{\footnotesize\par}
\par\end{raggedright}
\raggedright{}{\footnotesize{}c = TB-mBJ}{\footnotesize\par}
\end{table}

To obtain more band information, we analyze the total density of states
(TDOS) and projected density of electronic states (PDOS) using HSE06
xc functional. The compounds having a direct bandgap, the VBM and
CBM are both found at point $\Gamma$. The CBM consists of \textit{$s$}
orbitals of B-site cation (Si, Ge, Sn) and\textit{ $p$} orbitals
of halogen atom (Cl, Br, I), while the VBM is comprises only \textit{$p$}
orbitals of halogen atom. However, for Rb$_{2}$PtCl$_{6}$, the VBM
and CBM are located at a different point X, with a direct bandgap
of 4.35 eV. Here, contributions to the VBM and CBM come from Cl-\textit{$p$}
and Pt-\textit{$d$} orbitals. Figure \ref{fig:2} shows the TDOS
and PDOS of all the elements. Rb$_{2}$PtBr$_{6}$ and Rb$_{2}$PtI$_{6}$
exhibit indirect bandgaps, with the CBM and VBM located at the X and
$\Gamma$ points, respectively. Based on the PDOS calculations, in
Rb$_{2}$PtBr$_{6}$, VBM is mainly comprises Br-\textit{$p$} orbitals
and CBM is dominated by Br-\textit{$p$} orbitals with a small contribution
from Pt-\textit{$d$} orbitals. Similarly, in Rb$_{2}$PtI$_{6}$,
the VBM consists mainly of I-\textit{$p$} orbitals, and the CBM is
also primarily composed of I-\textit{$p$ }orbitals, with a minor
contribution from Pt-\textit{$d$} orbitals. Overall, the Rb$_{2}$BX$_{6}$
compounds demonstrate significant potential for optoelectronic applications,
with their bandgaps covering the infrared to ultraviolet range. Particularly,
Rb$_{2}$SnI$_{6}$ stands out as the most suitable one for photovoltaic
applications, owing to its direct bandgap of 1.16 eV, which falls
within the optimal range for solar cell performance.

To comprehend carrier transport behavior, we computed the effective
masses of charge carriers at the band edges of the aforementioned
systems. We derive the second-order derivative from the E-k\textbf{
}dispersion curve near the band edges using the equation $m^{\ast}=\hbar^{2}\left[\partial^{2}E(k)/\partial k^{2}\right]^{-1}$,
where $m^{\ast}$ represents the effective mass, $\hbar$ denotes
the reduced Planck\textquoteright s constant, $E(k)$ signifies the
energy eigenvalue, and $k$ stands for the wave vector. The effective
masses of electrons and holes are linked to the dispersion of the
CBM and VBM. Greater curvature in the CBM and VBM corresponds to lighter
effective masses. From the effective mass values presented in Table
\ref{tab:3}, we observe the electron effective masses tend to be
lower than their corresponding hole effective masses. This is primarily
attributed to the pronounced dispersive characteristics of the CBM
compared to the VBM, as evident from the band structures illustrated
in Figure \ref{fig:3}. The heavier hole masses, indicating Rb$_{2}$BX$_{6}$
VODPs are more suitable for \textit{n}-type semiconductors.

\begin{table}[H]
\caption{\label{tab:3}Carrier's effective mass, and reduced mass ($\mu^{*}$)
of Rb$\boldsymbol{_{2}}$BX$_{\boldsymbol{6}}$ (B = Si, Ge, Sn, Pt;
X = Cl, Br, I) VODPs calculated using G$_{0}$W$_{0}$@PBE method.
Here, $m_{e}$ and $m_{h}$ represents electron and hole effective
mass (where $m_{o}$ is the rest mass of the electron), respectively.
The bold values provided in parentheses are the effective mass and
reduced mass at direct band edge.}

\centering{}{\scriptsize{}}%
\begin{tabular}{cccccccccc}
\hline 
\multirow{1}{*}{{\scriptsize{}Configurations}} & {\scriptsize{}$m_{e}$ ($m_{0}$)} & {\scriptsize{}$m_{h}$ ($m_{0}$)} & {\scriptsize{}$\mu^{*}$ ($m_{0}$)} &  &  & {\scriptsize{}Configurations} & {\scriptsize{}$m_{e}$ ($m_{0}$)} & {\scriptsize{}$m_{h}$ ($m_{0}$)} & {\scriptsize{}$\mu^{*}$ ($m_{0}$)}\tabularnewline
\hline 
{\scriptsize{}Rb$_{2}$SiCl$_{6}$} & {\scriptsize{}0.641} & {\scriptsize{}1.527} & {\scriptsize{}0.451} &  &  & {\scriptsize{}Rb$_{2}$SnCl$_{6}$} & {\scriptsize{}0.580} & {\scriptsize{}1.392} & {\scriptsize{}0.409}\tabularnewline
{\scriptsize{}Rb$_{2}$SiBr$_{6}$} & {\scriptsize{}0.471} & {\scriptsize{}1.046} & {\scriptsize{}0.325} &  &  & {\scriptsize{}Rb$_{2}$SnBr$_{6}$} & {\scriptsize{}0.390} & {\scriptsize{}0.993} & {\scriptsize{}0.280}\tabularnewline
{\scriptsize{}Rb$_{2}$SiI$_{6}$} & {\scriptsize{}0.330} & {\scriptsize{}0.743} & {\scriptsize{}0.229} &  &  & {\scriptsize{}Rb$_{2}$SnI$_{6}$} & {\scriptsize{}0.186} & {\scriptsize{}0.682} & {\scriptsize{}0.146}\tabularnewline
{\scriptsize{}Rb$_{2}$GeCl$_{6}$} & {\scriptsize{}0.622} & {\scriptsize{}1.411} & {\scriptsize{}0.432} &  &  & {\scriptsize{}Rb$_{2}$PtCl$_{6}$} & {\scriptsize{}0.831} & {\scriptsize{}2.045} & {\scriptsize{}0.591}\tabularnewline
{\scriptsize{}Rb$_{2}$GeBr$_{6}$} & {\scriptsize{}0.383} & {\scriptsize{}0.987} & {\scriptsize{}0.276} &  &  & {\scriptsize{}Rb$_{2}$PtBr$_{6}$} & {\scriptsize{}0.559} & {\scriptsize{}0.997 (}\textbf{\scriptsize{}0.906}{\scriptsize{})} & {\scriptsize{}0.358 (}\textbf{\scriptsize{}0.346}{\scriptsize{})}\tabularnewline
{\scriptsize{}Rb$_{2}$GeI$_{6}$} & {\scriptsize{}0.268} & {\scriptsize{}0.687} & {\scriptsize{}0.193} &  &  & {\scriptsize{}Rb$_{2}$PtI$_{6}$} & {\scriptsize{}0.372} & {\scriptsize{}0.662 (}\textbf{\scriptsize{}0.681}{\scriptsize{})} & {\scriptsize{}0.238 (}\textbf{\scriptsize{}0.241}{\scriptsize{})}\tabularnewline
\hline 
\end{tabular}{\scriptsize\par}
\end{table}

\subsection{Optical Properties:}

Conducting a thorough investigation into the optical properties of
a material, including its dielectric function and absorption coefficient,
is fundamental for gaining a comprehensive understanding of its viability
for optoelectronic applications. In our contribution, the real {[}Re($\varepsilon$){]}
and imaginary {[}Im($\varepsilon$){]} part of the frequency-dependent
dielectric function are calculated to study the optical characteristics
of the chosen VODPs. The real part of the complex dielectric function
indicates how a material responds to an electric field, while the
imaginary part tells us how much of the light's energy is being consumed
as it passes through the material. To improve the accuracy of our
predictions, we employed the MBPT-based GW-BSE method, which explicitly
considers electron-hole interaction\citep{27,28}.

\begin{figure}[H]
\begin{centering}
\includegraphics[width=1\textwidth,height=1\textheight,keepaspectratio]{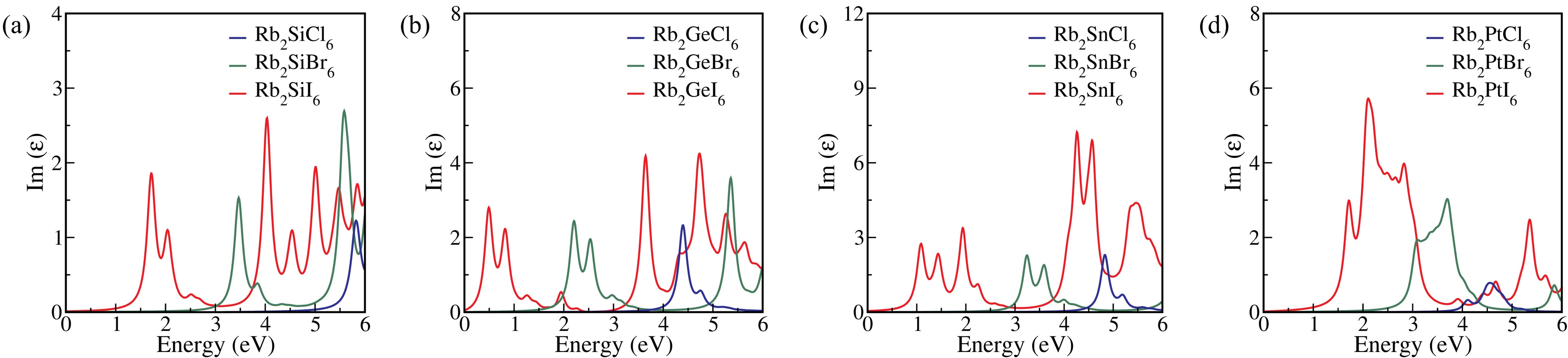}
\par\end{centering}
\caption{\label{fig:4}Imaginary {[}Im($\varepsilon$){]} part of the dielectric
function of (a) Rb$_{2}$SiX$_{6}$, (b) Rb$_{2}$GeX$_{6}$, (c)
Rb$_{2}$SnX$_{6}$, and (d) Rb$_{2}$PtX$_{6}$ VODPs, where X =
Cl, Br, and I, respectively, obtained using the BSE@G$_{0}$W$_{0}$@PBE
method.}
\end{figure}

The computed imaginary parts of the dielectric function are plotted
in Figure \ref{fig:4}. The first absorption peak appears in the range
0.490$-$5.822 eV. This primarily occurs due to the shift from X-$p$
orbitals at the upper edge of the valence band to the blending of
B-$s$ and X-$p$ orbitals hybridization states in the CBM. The optical
characteristics of Rb$_{2}$BX$_{6}$ compounds display comparable
patterns, with the initial peak positions varying based on distinct
bandgap values. As the bandgap increases, we notice that the edges
of the absorption spectra tend to shift towards the blue end. It's
interesting to note that materials with indirect bandgaps have absorption
edges significantly higher than their electronic bandgaps, while those
with direct bandgaps have absorption edges closer to their electronic
bandgaps. This suggests that in direct-band-gap materials, the first
optical transition occurs between the VBM and the CBM, unlike in indirect-band-gap
materials. Absorption-edge values derived from the BSE@G$_{0}$W$_{0}$@PBE
method fall within the 0.31-5.58 eV range, suggesting that these materials
absorb light spanning from the infrared to the ultraviolet region.
In conclusion, materials with absorption spanning the infrared to
ultraviolet regions present versatile optoelectronic properties, offering
potential for enhanced efficiency and performance across a wide range
of applications. This highlights the promise of Rb$_{2}$BX$_{6}$compounds
for various optoelectronic applications spanning diverse device types.

\subsection{Excitonic Properties:}

An exciton is a quasiparticle consisting of an electron and a hole
bound together by electrostatic forces. The electron and hole can
form a pair that behaves like a single, composite particle due to
their mutual attraction. The energy required to split the exciton
into an electron and a hole is defined as the exciton binding energy,
which is another crucial factor in photovoltaic applications. The
lower the exciton binding energy, the more easily the electron and
hole can be separated, which results in higher photoelectric conversion
efficiency. In our study, the exciton binding energy is calculated
using the Wannier-Mott (WM)\citep{74} formula given below:
\begin{center}
\begin{equation}
E_{B}=\left(\frac{\mu^{\ast}}{m_{0}\varepsilon_{\mathrm{eff}}^{2}}\right)R_{\infty},
\end{equation}
\par\end{center}

here $\mu^{\ast}$, $m_{0}$, $\varepsilon_{\mathrm{eff}}$, and $R_{\infty}$
stand for the reduced mass of charge carriers, rest mass of the electron,
effective dielectric constant and Rydberg constant respectively. $\mu^{\ast}$
is given by
\begin{center}
\begin{equation}
\frac{1}{\mu^{\ast}}=\frac{1}{m_{e}^{\ast}}+\frac{1}{m_{h}^{\ast}}
\end{equation}
\par\end{center}

In the case of Rb$_{2}$PtBr$_{6}$ and Rb$_{2}$PtI$_{6}$, which
have an indirect bandgap, we calculated the reduced mass ($\mu^{\ast}$)
of carriers by focusing on their lowest direct band edges (look at
Table \ref{tab:3}).

To calculate the exciton binding energy ($E_{B}$), we need $\varepsilon_{\mathrm{eff}}$.
If $E_{B}\ll\hbar\omega_{LO}$, lattice relaxation must be considered,
and an intermediate dielectric constant between $\varepsilon_{\infty}$
and $\varepsilon_{static}$ should be used. However, when $E_{B}\gg\hbar\omega_{LO}$,
as in our case, the ionic contribution becomes negligible, and the
effective dielectric constant ($\varepsilon_{\mathrm{eff}}$) approaches
$\varepsilon_{\infty}$, meaning the dielectric screening is dominated
by the electronic contribution\citep{79}.

We employed the density functional perturbation theory (DFPT) method
to estimate the ionic contribution to the dielectric function ($\varepsilon_{ion}$),
and the electronic contribution ($\varepsilon_{\infty}$) is estimated
using BSE@G$_{0}$W$_{0}$@PBE method (for details, see the Table
S4). The upper ($E_{Bu}$) and lower ($E_{Bl}$) bound to the exciton
binding energy is computed from the contribution of optical ($\varepsilon_{\infty}$)
and static ($\varepsilon_{static}$) dielectric constants, respectively.
The $E_{Bu}$ and $E_{Bl}$ values calculated using the WM method
are depicted in Table S4. $E_{Bu}$ and $E_{Bl}$ show a noticeable
decrease in double perovskite compounds when moving from Cl to I -compositions.
This decline is attributed to the higher dielectric constant of I-containing
compounds compared to their Cl and Br counterparts. Additionally,
the upper and lower limits of $E_{B}$ suggest that the electronic
contribution plays a more significant role than the ionic contribution
in the dielectric screening of these VODPs.

In addition to the WM model, we extended our analysis by calculating
the $E_{\ensuremath{B}}$ for Rb$_{2}$BX$_{6}$ systems using the
standard BSE@G$_{0}$W$_{0}$@PBE method\citep{75,76}. The formula
used for this calculation is $E_{B}=E_{g}^{dir}-E_{o}$, where $E_{g}^{dir}$
represents the GW (direct) bandgap, and $E_{o}$ corresponds to the
position of the first strong absorption peak from BSE calculation.
This approach provides a more comprehensive assessment of exciton
binding energy. In I-containing compounds, BSE@G$_{0}$W$_{0}$@PBE
and WM methods yield similar $E_{B}$ values, while the WM model predicts
stronger localized resonant excitons for Cl and Br-containing systems
ranging from 0.072 to 0.682 eV. Thus, the electronic contribution
dominates over the ionic contribution in dielectric screening for
I-containing VODPs.

{\scriptsize{}}
\begin{table}[H]
{\scriptsize{}\caption{\label{tab:4}Excitonic parameters of Rb$_{2}$BX$_{6}$ (B = Si,
Ge, Sn, Pt; X = Cl, Br, I) VODPs. Here, $E_{g}^{dir}$ is the direct
G$_{0}$W$_{0}$@PBE bandgap (in unit eV), $E_{o}$ represents the
exciton energy (in unit eV), $E_{B}$ represents the exciton binding
energy (in unit eV), $\Delta E_{B}^{ph}$ is the phonon screening
corrections of exciton binding energy (in unit meV), $(E_{B}+\Delta E_{B}^{ph})$
is the corrected values of exciton binding energy (in unit eV), $r_{exc}$
is the exciton radius (in unit nm), and $|\phi_{n}(0)|^{2}$ is the
probability of a wavefunction for electron-hole pair at zero separation
(in unit $m^{-3}$).}
}{\scriptsize\par}
\centering{}{\scriptsize{}}%
\begin{tabular}{ccccccccc}
\cline{1-1} \cline{3-9} \cline{4-9} \cline{5-9} \cline{6-9} \cline{7-9} \cline{8-9} \cline{9-9} 
\multirow{1}{*}{{\scriptsize{}Configurations}} &  & {\scriptsize{}$E_{g}^{dir}$ (eV)} & {\scriptsize{}$E_{o}$ (eV)} & {\scriptsize{}$E_{B}$ (eV)} & {\scriptsize{}$\Delta E_{B}^{ph}$ (meV)} & {\scriptsize{}$(E_{B}+\Delta E_{B}^{ph})$ (eV)} & {\scriptsize{}$r_{exc}$ (nm)} & {\scriptsize{}$|\phi_{n}(0)|^{2}$ ($10^{28}$$m^{-3}$)}\tabularnewline
\hline 
{\scriptsize{}Rb$_{2}$SiCl$_{6}$} &  & {\scriptsize{}6.122} & {\scriptsize{}5.822} & {\scriptsize{}0.300} & {\scriptsize{}-15.70} & {\scriptsize{}0.284} & {\scriptsize{}0.352} & {\scriptsize{}0.731}\tabularnewline
{\scriptsize{}Rb$_{2}$SiBr$_{6}$} &  & {\scriptsize{}3.664} & {\scriptsize{}3.467} & {\scriptsize{}0.197} & {\scriptsize{}-12.85} & {\scriptsize{}0.184} & {\scriptsize{}0.555} & {\scriptsize{}0.186}\tabularnewline
{\scriptsize{}Rb$_{2}$SiI$_{6}$} &  & {\scriptsize{}1.820} & {\scriptsize{}1.712} & {\scriptsize{}0.108} & {\scriptsize{}-10.30} & {\scriptsize{}0.098} & {\scriptsize{}1.107} & {\scriptsize{}0.023}\tabularnewline
{\scriptsize{}Rb$_{2}$GeCl$_{6}$} &  & {\scriptsize{}4.763} & {\scriptsize{}4.402} & {\scriptsize{}0.361} & {\scriptsize{}-13.22} & {\scriptsize{}0.348} & {\scriptsize{}0.298} & {\scriptsize{}1.208}\tabularnewline
{\scriptsize{}Rb$_{2}$GeBr$_{6}$} &  & {\scriptsize{}2.419} & {\scriptsize{}2.205} & {\scriptsize{}0.214} & {\scriptsize{}-9.23} & {\scriptsize{}0.205} & {\scriptsize{}0.594} & {\scriptsize{}0.152}\tabularnewline
{\scriptsize{}Rb$_{2}$GeI$_{6}$} &  & {\scriptsize{}0.561} & {\scriptsize{}0.490} & {\scriptsize{}0.071} & {\scriptsize{}-6.20} & {\scriptsize{}0.065} & {\scriptsize{}1.656} & {\scriptsize{}0.007}\tabularnewline
{\scriptsize{}Rb$_{2}$SnCl$_{6}$} &  & {\scriptsize{}5.245} & {\scriptsize{}4.825} & {\scriptsize{}0.420} & {\scriptsize{}-12.66} & {\scriptsize{}0.407} & {\scriptsize{}0.310} & {\scriptsize{}1.064}\tabularnewline
{\scriptsize{}Rb$_{2}$SnBr$_{6}$} &  & {\scriptsize{}3.379} & {\scriptsize{}3.247} & {\scriptsize{}0.132} & {\scriptsize{}-8.73} & {\scriptsize{}0.123} & {\scriptsize{}0.546} & {\scriptsize{}0.196}\tabularnewline
{\scriptsize{}Rb$_{2}$SnI$_{6}$} &  & {\scriptsize{}1.164} & {\scriptsize{}1.090} & {\scriptsize{}0.074} & {\scriptsize{}-6.70} & {\scriptsize{}0.067} & {\scriptsize{}1.399} & {\scriptsize{}0.012}\tabularnewline
{\scriptsize{}Rb$_{2}$PtCl$_{6}$} &  & {\scriptsize{}4.353} & {\scriptsize{}4.112} & {\scriptsize{}0.241} & {\scriptsize{}-10.35} & {\scriptsize{}0.231} & {\scriptsize{}0.251} & {\scriptsize{}2.022}\tabularnewline
{\scriptsize{}Rb$_{2}$PtBr$_{6}$} &  & {\scriptsize{}3.204} & {\scriptsize{}3.096} & {\scriptsize{}0.108} & {\scriptsize{}-6.12} & {\scriptsize{}0.102} & {\scriptsize{}0.547} & {\scriptsize{}0.194}\tabularnewline
{\scriptsize{}Rb$_{2}$PtI$_{6}$} &  & {\scriptsize{}1.786} & {\scriptsize{}1.710} & {\scriptsize{}0.076} & {\scriptsize{}-4.17} & {\scriptsize{}0.072} & {\scriptsize{}1.098} & {\scriptsize{}0.024}\tabularnewline
\hline 
\end{tabular}{\scriptsize\par}
\end{table}
{\scriptsize\par}

In a recent study, Filip and collaborators\citep{82} presented a
discussion on phonon screening into the calculation of exciton binding
energy. They assumed isotropic, parabolic electronic band dispersion
and considered four material parameters such as reduced effective
mass, static and optical dielectric constants, and the frequency of
the longitudinal optical phonon mode ($\omega_{LO}$). We determined
the characteristic frequency $\omega_{LO}$ using the athermal \textquotedbl B\textquotedbl{}
scheme proposed by Hellwarth et al.\citep{78}, which involves considering
the spectral average of infrared-active optical phonon modes (for
details, see the Supplemental Material). The correction due to phonon
screening is expressed as\citep{82}:
\begin{center}
\begin{equation}
\ensuremath{\ensuremath{\Delta}}E_{B}^{ph}=-2\omega_{LO}\left(1-\frac{\varepsilon_{\infty}}{\varepsilon_{static}}\right)\frac{\sqrt{1+\omega_{LO}/E_{B}}+3}{\left(1+\sqrt{1+\omega_{LO}/E_{B}}\right)^{3}}
\end{equation}
\par\end{center}

According to Table \ref{tab:4}, phonon screening leads to a decrease
in the exciton binding energy by 3.01 \% to 9.54 \%. While phonon
screening does reduce the exciton binding energy, the effect is not
particularly significant. This suggests that for VODPs, electronic
contribution plays a more dominant role in dielectric screening compared
to ionic or phonon contributions.

Further, using dielectric constant ($\varepsilon_{\mathrm{eff}}$)
and reduced mass of the charge carriers ($\mu^{\ast}$), we calculated
several excitonic properties such as excitonic radius ($\text{\ensuremath{r_{exc}}}$),
exciton lifetime ($\tau_{exc}$) and the probability of a wavefunction
($|\phi_{n}(0)|^{2}$) for electron-hole pair at zero separation.
$|\phi_{n}(0)|^{2}$ is determined using the following formula\citep{74},
\begin{center}
\begin{equation}
|\phi_{n}(0)|^{2}=\frac{1}{\pi(r_{exc})^{3}n^{3}}
\end{equation}
\par\end{center}

where $r_{exc}$ is the exciton radius, and n is the exciton energy
level. In order to evaluate $|\phi_{n}(0)|^{2}$, first we determined
excitonic radius $(r_{exc}$) by the given formula\citep{74}

\begin{equation}
r_{exc}=\frac{m_{0}}{\mu^{*}}\mathrm{\varepsilon_{eff}}n^{2}r_{Ry},
\end{equation}

here, $m_{0}$ is the rest mass of the electron, $\mu^{*}$ is the
reduced mass of charge carriers, $\mathrm{\varepsilon_{eff}}$ is
the effective dielectric constant, n is the exciton energy level (n=1
gives the smallest exciton radius) and $r_{Ry}$ is the Bohr radius
(0.0529 nm). In this scenario, the electronic contribution to the
dielectric function ($\varepsilon_{\infty}$) is regarded as the effective
value due to the insignificance of the ionic contribution to dielectric
screening. The exciton lifetime ($\tau_{exc}$) is inversely proportional
to the probability of a wavefunction ($|\phi_{n}(0)|^{2}$) for electron-hole
pair at zero separation (for details, for details, see Section IV
of the Supplemental Material). All the values of $|\phi_{n}(0)|^{2}$
are given in Table \ref{tab:4}, which suggests that exciton lifetime
keeps increasing while going from Cl to I-containing systems, and
I-containing compounds have longer exciton lifetime. A prolonged exciton
lifetime suggests reduced carrier recombination, leading to increased
conversion efficiency that is beneficial for photovoltaic applications.

\subsection{Polaronic Properties:}

Understanding polaronic parameters holds significant promise for advancing
material design and technology. The Fr\"ohlich mesoscopic model\citep{77}
reveals insights into the impact of longitudinal optical phonon modes
on carrier mobility, consequently improving device performance. According
to this model, the charge carriers moving through the lattice interact
with the polar optical phonons, and the interaction is given by the
dimensionless Fr\"ohlich parameter $\alpha$ as follows:
\begin{center}
\begin{equation}
\alpha=\left(\frac{1}{\varepsilon_{\infty}}-\frac{1}{\varepsilon_{static}}\right)\sqrt{\frac{R_{\infty}}{ch\omega_{LO}}}\sqrt{\frac{m^{*}}{m_{e}}}
\end{equation}
\par\end{center}

where, $\varepsilon_{\infty}$ and $\varepsilon_{static}$ represent
the electronic and static dielectric constants, respectively. Meanwhile,
$h$ stands for Planck's constant, $c$ denotes the speed of light,
and $R_{\infty}$ is Rydberg's constant. Notably, $\alpha\ll1$ typically
signifies weak electron (hole) -phonon coupling, whereas $\alpha>10$
denotes strong coupling\citep{77}. From the $\alpha$ values depicted
in Table \ref{tab:5}, we find an intermediate electron (hole)-phonon
coupling (1.66$-$9.68) for our configurations. Additionally, Debye
temperature ($\theta_{D}$) plays a key role in determining carrier-phonon
interactions. A lower $\theta_{D}$ relative to room temperature significantly
strengthens carrier-phonon coupling, enhancing polaronic effects.
In our study, most systems exhibit $\theta_{D}$ values lower than
room temperature, indicating significant carrier-phonon interactions
and increased polaronic contributions.

Using the values of $\alpha$, we can ascertain the decrease in the
quasiparticle (QP) gap resulting from polaron formation through the
following equation. Following equation describes the reduction in
QP energy for both electrons and holes\citep{79,80},
\begin{center}
\begin{equation}
E_{p}=(-\alpha-0.0123\alpha^{2})\hbar\omega_{LO}
\end{equation}
\par\end{center}

here, $E_{p}$ is the polaron energy. For our systems Rb$_{2}$SiCl$_{6}$,
Rb$_{2}$SiBr$_{6}$, Rb$_{2}$SiI$_{6}$, Rb$_{2}$GeCl$_{6}$, Rb$_{2}$GeBr$_{6}$,
Rb$_{2}$GeI$_{6}$, Rb$_{2}$SnCl$_{6}$, Rb$_{2}$SnBr$_{6}$, Rb$_{2}$SnI$_{6}$,
Rb$_{2}$PtCl$_{6}$, Rb$_{2}$PtBr$_{6}$, and Rb$_{2}$PtI$_{6}$,
the QP gap is lowered by 0.257, 0.173, 0.097, 0.306, 0.158, 0.052,
0.318, 0.184, 0.095, 0.26, 0.113, and 0.52 eV, respectively. Upon
comparison of these values with $E_{B}$ given in Table \ref{tab:4},
we can conclude that the charge-separated polaronic state is less
stable than the bound exciton except Rb$_{2}$SnBr$_{6}$, Rb$_{2}$SnI$_{6}$,
Rb$_{2}$PtCl$_{6}$, and Rb$_{2}$PtBr$_{6}$. Additionally, effective
polaron masses ($m_{p}$) are calculated by Feynman extended Frohlich\textquoteright s
polaron theory\citep{81}:
\begin{center}
\begin{equation}
m_{p}=m^{*}\left(1+\frac{\alpha}{6}+\frac{\alpha^{2}}{40}+...\right)
\end{equation}
\par\end{center}

where, $m^{*}$stands for effective mass calculated from bandstructure
calculations. This expression indicates that the polaron mass consistently
surpasses the effective mass obtained from band structure calculations,
thereby impacting charge carrier mobilities.

Next, we calculated Polaron mobilities ($\mu_{p}$) defined for electron
and hole with the help of the Hellwarth polaron model\citep{78}:
\begin{center}
\begin{equation}
\mu_{p}=\frac{\left(3\sqrt{\pi}e\right)}{2\pi c\omega_{LO}m^{*}\alpha}\frac{\sinh(\beta/2)}{\beta^{5/2}}\frac{w^{3}}{v^{3}}\frac{1}{K}
\end{equation}
\par\end{center}

where, $\beta=hc\omega_{LO}/k_{B}T$, $e$ represents the electronic
charge, $m^{\ast}$ denotes the effective mass of the charge carrier,
and $w$ and $v$ are associated with temperature-dependent variational
parameters (for details, see the Supplemental Material). $K$ is defined
as follows:
\begin{center}
\begin{equation}
K(a,b)=\int_{0}^{\infty}du\left[u^{2}+a^{2}-b\cos(vu)\right]^{-3/2}\cos(u)
\end{equation}
\par\end{center}

Here, $a^{2}$ and $b$ are evaluated as:
\begin{center}
\begin{equation}
a^{2}=(\beta/2)^{2}+\frac{(v^{2}-w^{2})}{w^{2}v}\beta\coth(\beta v/2)
\end{equation}
\par\end{center}

\begin{center}
\begin{equation}
b=\frac{v^{2}-w^{2}}{w^{2}v}\frac{\beta}{\sinh(\beta v/2)}
\end{equation}
\par\end{center}

The data presented in Table \ref{tab:5} reveals that systems having
iodine (I) exhibit higher polaron mobility compared to those containing
bromine (Br) and chlorine (Cl). This difference could be attributed
to variations in the electronegativity among different halogens. Hence,
it can be inferred that compounds containing iodine may serve as more
effective charge transport materials. The significant difference in
polaron mobility between electrons (3.33$-$85.11 cm$^{2}$V$^{-1}$s$^{-1}$)
and holes (0.34$-$17.98 cm$^{2}$V$^{-1}$s$^{-1}$) further supports
the characterization of Rb$_{2}$BX$_{6}$ as an n-type semiconductor.
In contrast, the electron polaron mobility for Cs-based VODPs ranges
from 42.15$-$71.12 cm$^{2}$V$^{-1}$s$^{-1}$\citep{86}, while
for Rb-based VODPs, it reaches up to 85.11 cm$^{2}$V$^{-1}$s$^{-1}$.
This difference highlights the superior charge transport capabilities
of our materials compared to previously studied Cs-based VODPs.

\begin{table}[H]
\caption{\label{tab:5}Polaron parameters corresponding to electrons ($e$)
and holes ($h$) in Rb$_{2}$BX$_{6}$ (B = Si, Ge, Sn, Pt; X = Cl,
Br, I) VODPs. Here, $\omega_{LO}$ represents the characteristic phonon
angular frequency (in unit THz), $\theta_{D}$ represents debye temperature
(in unit K), $\alpha$ represents Fr\"ohlich interaction parameter,
$m_{p}$ represents effective mass of the polaron ( in terms of $m^{*}$),
$E_{p}$ represents polaron energy (in unit meV), and $\mu_{p}$ represents
the polaron mobility (in unit cm$^{2}$V$^{-1}$s$^{-1}$), respectively.}

\centering{}{\scriptsize{}}%
\begin{tabular}{cccccccccccccc}
\toprule 
\multirow{2}{*}{{\scriptsize{}Configurations}} & \multirow{2}{*}{{\scriptsize{}$\omega_{LO}$ (THz)}} & \multirow{2}{*}{{\scriptsize{}$\theta_{D}$ (K)}} & \multicolumn{2}{c}{{\scriptsize{}$\alpha$}} &  & \multicolumn{2}{c}{{\scriptsize{}$m_{p}/m^{*}$}} &  & \multicolumn{2}{c}{{\scriptsize{}$E_{p}$ (meV)}} &  & \multicolumn{2}{c}{{\scriptsize{}$\mu_{p}$ (cm$^{2}$V$^{-1}$s$^{-1}$)}}\tabularnewline
\cmidrule{4-5} \cmidrule{5-5} \cmidrule{7-8} \cmidrule{8-8} \cmidrule{10-11} \cmidrule{11-11} \cmidrule{13-14} \cmidrule{14-14} 
 &  &  & {\scriptsize{}$e$} & {\scriptsize{}$h$} &  & {\scriptsize{}$e$} & {\scriptsize{}$h$} &  & {\scriptsize{}$e$} & {\scriptsize{}$h$} &  & {\scriptsize{}$e$} & {\scriptsize{}$h$}\tabularnewline
\midrule
{\scriptsize{}Rb$_{2}$SiCl$_{6}$} & {\scriptsize{}7.06} & {\scriptsize{}339} & {\scriptsize{}3.28} & {\scriptsize{}5.06} &  & {\scriptsize{}1.82} & {\scriptsize{}2.48} &  & {\scriptsize{}99.77} & {\scriptsize{}157.15} &  & {\scriptsize{}9.01} & {\scriptsize{}1.53}\tabularnewline
{\scriptsize{}Rb$_{2}$SiBr$_{6}$} & {\scriptsize{}5.76} & {\scriptsize{}277} & {\scriptsize{}2.78} & {\scriptsize{}4.15} &  & {\scriptsize{}1.66} & {\scriptsize{}2.12} &  & {\scriptsize{}68.58} & {\scriptsize{}104.05} &  & {\scriptsize{}17.40} & {\scriptsize{}3.85}\tabularnewline
{\scriptsize{}Rb$_{2}$SiI$_{6}$} & {\scriptsize{}4.31} & {\scriptsize{}207} & {\scriptsize{}2.10} & {\scriptsize{}3.16} &  & {\scriptsize{}1.46} & {\scriptsize{}1.78} &  & {\scriptsize{}38.45} & {\scriptsize{}58.59} &  & {\scriptsize{}42.25} & {\scriptsize{}10.29}\tabularnewline
{\scriptsize{}Rb$_{2}$GeCl$_{6}$} & {\scriptsize{}5.12} & {\scriptsize{}246} & {\scriptsize{}5.31} & {\scriptsize{}8.00} &  & {\scriptsize{}2.59} & {\scriptsize{}3.93} &  & {\scriptsize{}119.94} & {\scriptsize{}186.32} &  & {\scriptsize{}4.07} & {\scriptsize{}0.56}\tabularnewline
{\scriptsize{}Rb$_{2}$GeBr$_{6}$} & {\scriptsize{}3.74} & {\scriptsize{}180} & {\scriptsize{}3.68} & {\scriptsize{}5.92} &  & {\scriptsize{}1.95} & {\scriptsize{}2.86} &  & {\scriptsize{}59.58} & {\scriptsize{}98.37} &  & {\scriptsize{}16.87} & {\scriptsize{}2.57}\tabularnewline
{\scriptsize{}Rb$_{2}$GeI$_{6}$} & {\scriptsize{}2.88} & {\scriptsize{}138} & {\scriptsize{}1.66} & {\scriptsize{}2.66} &  & {\scriptsize{}1.35} & {\scriptsize{}1.62} &  & {\scriptsize{}20.20} & {\scriptsize{}32.76} &  & {\scriptsize{}85.11} & {\scriptsize{}17.98}\tabularnewline
{\scriptsize{}Rb$_{2}$SnCl$_{6}$} & {\scriptsize{}4.37} & {\scriptsize{}210} & {\scriptsize{}6.25} & {\scriptsize{}9.68} &  & {\scriptsize{}3.02} & {\scriptsize{}4.96} &  & {\scriptsize{}121.80} & {\scriptsize{}196.04} &  & {\scriptsize{}3.33} & {\scriptsize{}0.34}\tabularnewline
{\scriptsize{}Rb$_{2}$SnBr$_{6}$} & {\scriptsize{}3.09} & {\scriptsize{}148} & {\scriptsize{}5.09} & {\scriptsize{}8.13} &  & {\scriptsize{}2.50} & {\scriptsize{}4.01} &  & {\scriptsize{}69.21} & {\scriptsize{}114.44} &  & {\scriptsize{}10.54} & {\scriptsize{}1.36}\tabularnewline
{\scriptsize{}Rb$_{2}$SnI$_{6}$} & {\scriptsize{}2.29} & {\scriptsize{}110} & {\scriptsize{}3.23} & {\scriptsize{}6.18} &  & {\scriptsize{}1.80} & {\scriptsize{}2.99} &  & {\scriptsize{}31.85} & {\scriptsize{}63.06} &  & {\scriptsize{}57.81} & {\scriptsize{}5.28}\tabularnewline
{\scriptsize{}Rb$_{2}$PtCl$_{6}$} & {\scriptsize{}4.40} & {\scriptsize{}211} & {\scriptsize{}5.11} & {\scriptsize{}8.02} &  & {\scriptsize{}2.50} & {\scriptsize{}3.94} &  & {\scriptsize{}98.96} & {\scriptsize{}160.55} &  & {\scriptsize{}3.72} & {\scriptsize{}0.45}\tabularnewline
{\scriptsize{}Rb$_{2}$PtBr$_{6}$} & {\scriptsize{}2.92} & {\scriptsize{}140} & {\scriptsize{}3.79} & {\scriptsize{}5.07} &  & {\scriptsize{}1.99} & {\scriptsize{}2.49} &  & {\scriptsize{}47.97} & {\scriptsize{}65.13} &  & {\scriptsize{}12.91} & {\scriptsize{}4.35}\tabularnewline
{\scriptsize{}Rb$_{2}$PtI$_{6}$} & {\scriptsize{}2.11} & {\scriptsize{}101} & {\scriptsize{}2.46} & {\scriptsize{}3.28} &  & {\scriptsize{}1.56} & {\scriptsize{}1.82} &  & {\scriptsize{}22.15} & {\scriptsize{}29.82} &  & {\scriptsize{}44.04} & {\scriptsize{}16.74}\tabularnewline
\bottomrule
\end{tabular}{\scriptsize\par}
\end{table}

\section{Conclusions:}

In summary, our investigation into Rb$_{2}$BX$_{6}$ systems unveils
their adherence to the standard cubic double-perovskite lattice {[}space
group Fm$\bar{3}$m (225){]}, along with maintaining ideal tolerance
and octahedral factors, as per the crystallographic stability criterion.
Additionally, the negative formation energies indicate the phase stability
of these systems. Moreover, our analysis demonstrates their mechanical
stability and elastic characteristics, revealing their flexible and
ductile nature, accompanied by elastic anisotropy. Employing GW-based
electronic band-structure calculations, we anticipate a broad tunability
of the bandgap within the range of 0.56 to 6.12 eV for direct bandgap
materials and 1.76 to 3.18 eV for indirect bandgap materials. Notably,
incorporating halogen atoms results in varied band-gap values, with
Cl-containing compounds exhibiting larger bandgaps than their Br and
I-containing counterparts. Furthermore, our findings indicate the
absorption-edge values fall within the range of 0.31$-$5.58 eV, enabling
efficient light absorption from infrared to ultraviolet wavelengths.
Remarkably, these materials exhibit low carrier effective masses,
suggesting exceptional charge-carrier mobility and lighter electron
masses, indicating these materials are more suitable for n-type semiconductors.
Excitonic properties indicates low to moderate exciton-binding energies
and varying exciton lifetimes, suggesting the potential for higher
quantum yield and conversion efficiency. Finally, investigating polaronic
properties unveils enhanced polaron mobility (ranging from 3.33 to
85.11 cm$^{2}$V$^{-1}$s$^{-1}$ for electrons) than the previously
reported VODPs materials. Additionally, a comparison between polaron
energy and exciton binding energy indicates a less stable state for
charge-separated polarons except Rb$_{2}$SnBr$_{6}$, Rb$_{2}$SnI$_{6}$,
Rb$_{2}$PtCl$_{6}$, and Rb$_{2}$PtBr$_{6}$. Collectively, our
study underscores the potential of Rb$_{2}$BX$_{6}$ VODP compounds
in various optoelectronic applications, owing to their versatile properties.
\begin{acknowledgments}
S.A. would like to acknowledge the Council of Scientific and Industrial
Research (CSIR), Government of India {[}Grant No. 09/1128(11453)/2021-EMR-I{]}
for Senior Research Fellowship. A.C. would also like to acknowledge
the Shiv Nadar Foundation for financial support. The authors acknowledge
the High Performance Computing Cluster (HPCC) \textquoteleft Magus\textquoteright{}
at Shiv Nadar Institution of Eminence for providing computational
resources that have contributed to the research results reported within
this paper.
\end{acknowledgments}

\bibliographystyle{apsrev4-2}
\bibliography{refs}

\begin{thebibliography}{68}%
\makeatletter
\providecommand \@ifxundefined [1]{%
 \@ifx{#1\undefined}
}%
\providecommand \@ifnum [1]{%
 \ifnum #1\expandafter \@firstoftwo
 \else \expandafter \@secondoftwo
 \fi
}%
\providecommand \@ifx [1]{%
 \ifx #1\expandafter \@firstoftwo
 \else \expandafter \@secondoftwo
 \fi
}%
\providecommand \natexlab [1]{#1}%
\providecommand \enquote  [1]{``#1''}%
\providecommand \bibnamefont  [1]{#1}%
\providecommand \bibfnamefont [1]{#1}%
\providecommand \citenamefont [1]{#1}%
\providecommand \href@noop [0]{\@secondoftwo}%
\providecommand \href [0]{\begingroup \@sanitize@url \@href}%
\providecommand \@href[1]{\@@startlink{#1}\@@href}%
\providecommand \@@href[1]{\endgroup#1\@@endlink}%
\providecommand \@sanitize@url [0]{\catcode `\\12\catcode `\$12\catcode `\&12\catcode `\#12\catcode `\^12\catcode `\_12\catcode `\%12\relax}%
\providecommand \@@startlink[1]{}%
\providecommand \@@endlink[0]{}%
\providecommand \url  [0]{\begingroup\@sanitize@url \@url }%
\providecommand \@url [1]{\endgroup\@href {#1}{\urlprefix }}%
\providecommand \urlprefix  [0]{URL }%
\providecommand \Eprint [0]{\href }%
\providecommand \doibase [0]{https://doi.org/}%
\providecommand \selectlanguage [0]{\@gobble}%
\providecommand \bibinfo  [0]{\@secondoftwo}%
\providecommand \bibfield  [0]{\@secondoftwo}%
\providecommand \translation [1]{[#1]}%
\providecommand \BibitemOpen [0]{}%
\providecommand \bibitemStop [0]{}%
\providecommand \bibitemNoStop [0]{.\EOS\space}%
\providecommand \EOS [0]{\spacefactor3000\relax}%
\providecommand \BibitemShut  [1]{\csname bibitem#1\endcsname}%
\let\auto@bib@innerbib\@empty
\bibitem [{\citenamefont {Hussain}\ \emph {et~al.}(2020)\citenamefont {Hussain}, \citenamefont {Rashid}, \citenamefont {Ali}, \citenamefont {Bhopal},\ and\ \citenamefont {Bhatti}}]{21}%
  \BibitemOpen
  \bibfield  {author} {\bibinfo {author} {\bibfnamefont {M.}~\bibnamefont {Hussain}}, \bibinfo {author} {\bibfnamefont {M.}~\bibnamefont {Rashid}}, \bibinfo {author} {\bibfnamefont {A.}~\bibnamefont {Ali}}, \bibinfo {author} {\bibfnamefont {M.~F.}\ \bibnamefont {Bhopal}},\ and\ \bibinfo {author} {\bibfnamefont {A.}~\bibnamefont {Bhatti}},\ }\href {https://doi.org/https://doi.org/10.1016/j.ceramint.2020.05.235} {\bibfield  {journal} {\bibinfo  {journal} {Ceram. Int.}\ }\textbf {\bibinfo {volume} {46}},\ \bibinfo {pages} {21378} (\bibinfo {year} {2020})}\BibitemShut {NoStop}%
\bibitem [{\citenamefont {Bush}\ \emph {et~al.}(2017)\citenamefont {Bush}, \citenamefont {Palmstrom}, \citenamefont {Yu}, \citenamefont {Boccard}, \citenamefont {Cheacharoen}, \citenamefont {Mailoa}, \citenamefont {McMeekin}, \citenamefont {Hoye}, \citenamefont {Bailie}, \citenamefont {Leijtens}, \citenamefont {Peters}, \citenamefont {Minichetti}, \citenamefont {Rolston}, \citenamefont {Prasanna}, \citenamefont {Sofia}, \citenamefont {Harwood}, \citenamefont {Ma}, \citenamefont {Moghadam}, \citenamefont {Snaith}, \citenamefont {Buonassisi}, \citenamefont {Holman}, \citenamefont {Bent},\ and\ \citenamefont {McGehee}}]{92}%
  \BibitemOpen
  \bibfield  {author} {\bibinfo {author} {\bibfnamefont {K.~A.}\ \bibnamefont {Bush}}, \bibinfo {author} {\bibfnamefont {A.~F.}\ \bibnamefont {Palmstrom}}, \bibinfo {author} {\bibfnamefont {Z.~J.}\ \bibnamefont {Yu}}, \bibinfo {author} {\bibfnamefont {M.}~\bibnamefont {Boccard}}, \bibinfo {author} {\bibfnamefont {R.}~\bibnamefont {Cheacharoen}}, \bibinfo {author} {\bibfnamefont {J.~P.}\ \bibnamefont {Mailoa}}, \bibinfo {author} {\bibfnamefont {D.~P.}\ \bibnamefont {McMeekin}}, \bibinfo {author} {\bibfnamefont {R.~L.~Z.}\ \bibnamefont {Hoye}}, \bibinfo {author} {\bibfnamefont {C.~D.}\ \bibnamefont {Bailie}}, \bibinfo {author} {\bibfnamefont {T.}~\bibnamefont {Leijtens}}, \bibinfo {author} {\bibfnamefont {I.~M.}\ \bibnamefont {Peters}}, \bibinfo {author} {\bibfnamefont {M.~C.}\ \bibnamefont {Minichetti}}, \bibinfo {author} {\bibfnamefont {N.}~\bibnamefont {Rolston}}, \bibinfo {author} {\bibfnamefont {R.}~\bibnamefont {Prasanna}}, \bibinfo {author} {\bibfnamefont {S.}~\bibnamefont {Sofia}}, \bibinfo {author}
  {\bibfnamefont {D.}~\bibnamefont {Harwood}}, \bibinfo {author} {\bibfnamefont {W.}~\bibnamefont {Ma}}, \bibinfo {author} {\bibfnamefont {F.}~\bibnamefont {Moghadam}}, \bibinfo {author} {\bibfnamefont {H.~J.}\ \bibnamefont {Snaith}}, \bibinfo {author} {\bibfnamefont {T.}~\bibnamefont {Buonassisi}}, \bibinfo {author} {\bibfnamefont {Z.~C.}\ \bibnamefont {Holman}}, \bibinfo {author} {\bibfnamefont {S.~F.}\ \bibnamefont {Bent}},\ and\ \bibinfo {author} {\bibfnamefont {M.~D.}\ \bibnamefont {McGehee}},\ }\href {https://doi.org/10.1038/nenergy.2017.9} {\bibfield  {journal} {\bibinfo  {journal} {Nat. Energy}\ }\textbf {\bibinfo {volume} {2}},\ \bibinfo {pages} {17009} (\bibinfo {year} {2017})}\BibitemShut {NoStop}%
\bibitem [{\citenamefont {Jana}\ \emph {et~al.}(2019)\citenamefont {Jana}, \citenamefont {Janke}, \citenamefont {Dirkes}, \citenamefont {Dovletgeldi}, \citenamefont {Liu}, \citenamefont {Qin}, \citenamefont {Gundogdu}, \citenamefont {You}, \citenamefont {Blum},\ and\ \citenamefont {Mitzi}}]{49}%
  \BibitemOpen
  \bibfield  {author} {\bibinfo {author} {\bibfnamefont {M.~K.}\ \bibnamefont {Jana}}, \bibinfo {author} {\bibfnamefont {S.~M.}\ \bibnamefont {Janke}}, \bibinfo {author} {\bibfnamefont {D.~J.}\ \bibnamefont {Dirkes}}, \bibinfo {author} {\bibfnamefont {S.}~\bibnamefont {Dovletgeldi}}, \bibinfo {author} {\bibfnamefont {C.}~\bibnamefont {Liu}}, \bibinfo {author} {\bibfnamefont {X.}~\bibnamefont {Qin}}, \bibinfo {author} {\bibfnamefont {K.}~\bibnamefont {Gundogdu}}, \bibinfo {author} {\bibfnamefont {W.}~\bibnamefont {You}}, \bibinfo {author} {\bibfnamefont {V.}~\bibnamefont {Blum}},\ and\ \bibinfo {author} {\bibfnamefont {D.~B.}\ \bibnamefont {Mitzi}},\ }\href {https://doi.org/10.1021/jacs.9b02909} {\bibfield  {journal} {\bibinfo  {journal} {J. Am. Chem. Soc.}\ }\textbf {\bibinfo {volume} {141}},\ \bibinfo {pages} {7955} (\bibinfo {year} {2019})},\ \bibinfo {note} {pMID: 31017429},\ \Eprint {https://arxiv.org/abs/https://doi.org/10.1021/jacs.9b02909} {https://doi.org/10.1021/jacs.9b02909} \BibitemShut {NoStop}%
\bibitem [{\citenamefont {Adhikari}\ and\ \citenamefont {Johari}(2023)}]{93}%
  \BibitemOpen
  \bibfield  {author} {\bibinfo {author} {\bibfnamefont {S.}~\bibnamefont {Adhikari}}\ and\ \bibinfo {author} {\bibfnamefont {P.}~\bibnamefont {Johari}},\ }\href {https://doi.org/10.1103/PhysRevMaterials.7.075401} {\bibfield  {journal} {\bibinfo  {journal} {Phys. Rev. Mater.}\ }\textbf {\bibinfo {volume} {7}},\ \bibinfo {pages} {075401} (\bibinfo {year} {2023})}\BibitemShut {NoStop}%
\bibitem [{\citenamefont {Nazir}\ \emph {et~al.}(2024)\citenamefont {Nazir}, \citenamefont {Maqsood}, \citenamefont {Noor}, \citenamefont {Mumtaz},\ and\ \citenamefont {Moussa}}]{94}%
  \BibitemOpen
  \bibfield  {author} {\bibinfo {author} {\bibfnamefont {S.}~\bibnamefont {Nazir}}, \bibinfo {author} {\bibfnamefont {S.}~\bibnamefont {Maqsood}}, \bibinfo {author} {\bibfnamefont {N.}~\bibnamefont {Noor}}, \bibinfo {author} {\bibfnamefont {S.}~\bibnamefont {Mumtaz}},\ and\ \bibinfo {author} {\bibfnamefont {I.~M.}\ \bibnamefont {Moussa}},\ }\href {https://doi.org/https://doi.org/10.1016/j.poly.2024.117003} {\bibfield  {journal} {\bibinfo  {journal} {Polyhedron}\ }\textbf {\bibinfo {volume} {256}},\ \bibinfo {pages} {117003} (\bibinfo {year} {2024})}\BibitemShut {NoStop}%
\bibitem [{\citenamefont {Zhang}\ \emph {et~al.}(2016)\citenamefont {Zhang}, \citenamefont {Eperon},\ and\ \citenamefont {Snaith}}]{70}%
  \BibitemOpen
  \bibfield  {author} {\bibinfo {author} {\bibfnamefont {W.}~\bibnamefont {Zhang}}, \bibinfo {author} {\bibfnamefont {G.~E.}\ \bibnamefont {Eperon}},\ and\ \bibinfo {author} {\bibfnamefont {H.~J.}\ \bibnamefont {Snaith}},\ }\href {https://doi.org/10.1038/nenergy.2016.48} {\bibfield  {journal} {\bibinfo  {journal} {Nat. Energys}\ }\textbf {\bibinfo {volume} {1}},\ \bibinfo {pages} {16048} (\bibinfo {year} {2016})}\BibitemShut {NoStop}%
\bibitem [{\citenamefont {Yin}\ \emph {et~al.}(2014)\citenamefont {Yin}, \citenamefont {Shi},\ and\ \citenamefont {Yan}}]{71}%
  \BibitemOpen
  \bibfield  {author} {\bibinfo {author} {\bibfnamefont {W.-J.}\ \bibnamefont {Yin}}, \bibinfo {author} {\bibfnamefont {T.}~\bibnamefont {Shi}},\ and\ \bibinfo {author} {\bibfnamefont {Y.}~\bibnamefont {Yan}},\ }\href {https://doi.org/https://doi.org/10.1002/adma.201306281} {\bibfield  {journal} {\bibinfo  {journal} {Adv. Mater.}\ }\textbf {\bibinfo {volume} {26}},\ \bibinfo {pages} {4653} (\bibinfo {year} {2014})},\ \Eprint {https://arxiv.org/abs/https://onlinelibrary.wiley.com/doi/pdf/10.1002/adma.201306281} {https://onlinelibrary.wiley.com/doi/pdf/10.1002/adma.201306281} \BibitemShut {NoStop}%
\bibitem [{\citenamefont {Dotsenko}\ \emph {et~al.}(2019)\citenamefont {Dotsenko}, \citenamefont {Vovna}, \citenamefont {Korochentsev}, \citenamefont {Mirochnik}, \citenamefont {Shcheka}, \citenamefont {Sedakova},\ and\ \citenamefont {Sergienko}}]{4}%
  \BibitemOpen
  \bibfield  {author} {\bibinfo {author} {\bibfnamefont {A.~A.}\ \bibnamefont {Dotsenko}}, \bibinfo {author} {\bibfnamefont {V.~I.}\ \bibnamefont {Vovna}}, \bibinfo {author} {\bibfnamefont {V.~V.}\ \bibnamefont {Korochentsev}}, \bibinfo {author} {\bibfnamefont {A.~G.}\ \bibnamefont {Mirochnik}}, \bibinfo {author} {\bibfnamefont {O.~L.}\ \bibnamefont {Shcheka}}, \bibinfo {author} {\bibfnamefont {T.~V.}\ \bibnamefont {Sedakova}},\ and\ \bibinfo {author} {\bibfnamefont {V.~I.}\ \bibnamefont {Sergienko}},\ }\href {https://doi.org/10.1021/acs.inorgchem.9b00250} {\bibfield  {journal} {\bibinfo  {journal} {Inorg. Chem.}\ }\textbf {\bibinfo {volume} {58}},\ \bibinfo {pages} {6796} (\bibinfo {year} {2019})},\ \Eprint {https://arxiv.org/abs/https://doi.org/10.1021/acs.inorgchem.9b00250} {https://doi.org/10.1021/acs.inorgchem.9b00250} \BibitemShut {NoStop}%
\bibitem [{\citenamefont {Maughan}\ \emph {et~al.}(2016)\citenamefont {Maughan}, \citenamefont {Ganose}, \citenamefont {Bordelon}, \citenamefont {Miller}, \citenamefont {Scanlon},\ and\ \citenamefont {Neilson}}]{2}%
  \BibitemOpen
  \bibfield  {author} {\bibinfo {author} {\bibfnamefont {A.~E.}\ \bibnamefont {Maughan}}, \bibinfo {author} {\bibfnamefont {A.~M.}\ \bibnamefont {Ganose}}, \bibinfo {author} {\bibfnamefont {M.~M.}\ \bibnamefont {Bordelon}}, \bibinfo {author} {\bibfnamefont {E.~M.}\ \bibnamefont {Miller}}, \bibinfo {author} {\bibfnamefont {D.~O.}\ \bibnamefont {Scanlon}},\ and\ \bibinfo {author} {\bibfnamefont {J.~R.}\ \bibnamefont {Neilson}},\ }\href {https://doi.org/10.1021/jacs.6b03207} {\bibfield  {journal} {\bibinfo  {journal} {J. Am. Chem. Soc.}\ }\textbf {\bibinfo {volume} {138}},\ \bibinfo {pages} {8453} (\bibinfo {year} {2016})},\ \bibinfo {note} {pMID: 27284638},\ \Eprint {https://arxiv.org/abs/https://doi.org/10.1021/jacs.6b03207} {https://doi.org/10.1021/jacs.6b03207} \BibitemShut {NoStop}%
\bibitem [{\citenamefont {Karim}\ \emph {et~al.}(2019)\citenamefont {Karim}, \citenamefont {Ganose}, \citenamefont {Pieters}, \citenamefont {Winnie~Leung}, \citenamefont {Wade}, \citenamefont {Zhang}, \citenamefont {Scanlon},\ and\ \citenamefont {Palgrave}}]{7}%
  \BibitemOpen
  \bibfield  {author} {\bibinfo {author} {\bibfnamefont {M.~M.~S.}\ \bibnamefont {Karim}}, \bibinfo {author} {\bibfnamefont {A.~M.}\ \bibnamefont {Ganose}}, \bibinfo {author} {\bibfnamefont {L.}~\bibnamefont {Pieters}}, \bibinfo {author} {\bibfnamefont {W.~W.}\ \bibnamefont {Winnie~Leung}}, \bibinfo {author} {\bibfnamefont {J.}~\bibnamefont {Wade}}, \bibinfo {author} {\bibfnamefont {L.}~\bibnamefont {Zhang}}, \bibinfo {author} {\bibfnamefont {D.~O.}\ \bibnamefont {Scanlon}},\ and\ \bibinfo {author} {\bibfnamefont {R.~G.}\ \bibnamefont {Palgrave}},\ }\href {https://doi.org/10.1021/acs.chemmater.9b03267} {\bibfield  {journal} {\bibinfo  {journal} {Chem. Mater.}\ }\textbf {\bibinfo {volume} {31}},\ \bibinfo {pages} {9430} (\bibinfo {year} {2019})},\ \bibinfo {note} {pMID: 32116409},\ \Eprint {https://arxiv.org/abs/https://doi.org/10.1021/acs.chemmater.9b03267} {https://doi.org/10.1021/acs.chemmater.9b03267} \BibitemShut {NoStop}%
\bibitem [{\citenamefont {Travis}\ \emph {et~al.}(2016)\citenamefont {Travis}, \citenamefont {Glover}, \citenamefont {Bronstein}, \citenamefont {Scanlon},\ and\ \citenamefont {Palgrave}}]{8}%
  \BibitemOpen
  \bibfield  {author} {\bibinfo {author} {\bibfnamefont {W.}~\bibnamefont {Travis}}, \bibinfo {author} {\bibfnamefont {E.~N.~K.}\ \bibnamefont {Glover}}, \bibinfo {author} {\bibfnamefont {H.}~\bibnamefont {Bronstein}}, \bibinfo {author} {\bibfnamefont {D.~O.}\ \bibnamefont {Scanlon}},\ and\ \bibinfo {author} {\bibfnamefont {R.~G.}\ \bibnamefont {Palgrave}},\ }\href {https://doi.org/10.1039/C5SC04845A} {\bibfield  {journal} {\bibinfo  {journal} {Chem. Sci.}\ }\textbf {\bibinfo {volume} {7}},\ \bibinfo {pages} {4548} (\bibinfo {year} {2016})}\BibitemShut {NoStop}%
\bibitem [{\citenamefont {Faizan}\ \emph {et~al.}(2021{\natexlab{a}})\citenamefont {Faizan}, \citenamefont {Xie}, \citenamefont {Murtaza}, \citenamefont {Echeverr\'{i}a-Arrondo}, \citenamefont {Alshahrani}, \citenamefont {Bhamu}, \citenamefont {Laref}, \citenamefont {Mora-Ser\'{o}},\ and\ \citenamefont {Haidar~Khan}}]{9}%
  \BibitemOpen
  \bibfield  {author} {\bibinfo {author} {\bibfnamefont {M.}~\bibnamefont {Faizan}}, \bibinfo {author} {\bibfnamefont {J.}~\bibnamefont {Xie}}, \bibinfo {author} {\bibfnamefont {G.}~\bibnamefont {Murtaza}}, \bibinfo {author} {\bibfnamefont {C.}~\bibnamefont {Echeverr\'{i}a-Arrondo}}, \bibinfo {author} {\bibfnamefont {T.}~\bibnamefont {Alshahrani}}, \bibinfo {author} {\bibfnamefont {K.~C.}\ \bibnamefont {Bhamu}}, \bibinfo {author} {\bibfnamefont {A.}~\bibnamefont {Laref}}, \bibinfo {author} {\bibfnamefont {I.}~\bibnamefont {Mora-Ser\'{o}}},\ and\ \bibinfo {author} {\bibfnamefont {S.}~\bibnamefont {Haidar~Khan}},\ }\href {https://doi.org/10.1039/D0CP05827K} {\bibfield  {journal} {\bibinfo  {journal} {Phys. Chem. Chem. Phys.}\ }\textbf {\bibinfo {volume} {23}},\ \bibinfo {pages} {4646} (\bibinfo {year} {2021}{\natexlab{a}})}\BibitemShut {NoStop}%
\bibitem [{\citenamefont {Faizan}\ \emph {et~al.}(2021{\natexlab{b}})\citenamefont {Faizan}, \citenamefont {Bhamu}, \citenamefont {Murtaza}, \citenamefont {He}, \citenamefont {Kulhari}, \citenamefont {AL-Anazy},\ and\ \citenamefont {Khan}}]{61}%
  \BibitemOpen
  \bibfield  {author} {\bibinfo {author} {\bibfnamefont {M.}~\bibnamefont {Faizan}}, \bibinfo {author} {\bibfnamefont {K.~C.}\ \bibnamefont {Bhamu}}, \bibinfo {author} {\bibfnamefont {G.}~\bibnamefont {Murtaza}}, \bibinfo {author} {\bibfnamefont {X.}~\bibnamefont {He}}, \bibinfo {author} {\bibfnamefont {N.}~\bibnamefont {Kulhari}}, \bibinfo {author} {\bibfnamefont {M.~M.}\ \bibnamefont {AL-Anazy}},\ and\ \bibinfo {author} {\bibfnamefont {S.~H.}\ \bibnamefont {Khan}},\ }\href {https://doi.org/10.1038/s41598-021-86145-x} {\bibfield  {journal} {\bibinfo  {journal} {Sci. Rep.}\ }\textbf {\bibinfo {volume} {11}},\ \bibinfo {pages} {6965} (\bibinfo {year} {2021}{\natexlab{b}})}\BibitemShut {NoStop}%
\bibitem [{\citenamefont {Zhao}\ \emph {et~al.}(2021)\citenamefont {Zhao}, \citenamefont {Wei}, \citenamefont {Tang}, \citenamefont {Gao}, \citenamefont {Xie}, \citenamefont {Lu},\ and\ \citenamefont {Tang}}]{11}%
  \BibitemOpen
  \bibfield  {author} {\bibinfo {author} {\bibfnamefont {X.-H.}\ \bibnamefont {Zhao}}, \bibinfo {author} {\bibfnamefont {X.-N.}\ \bibnamefont {Wei}}, \bibinfo {author} {\bibfnamefont {T.-Y.}\ \bibnamefont {Tang}}, \bibinfo {author} {\bibfnamefont {L.-K.}\ \bibnamefont {Gao}}, \bibinfo {author} {\bibfnamefont {Q.}~\bibnamefont {Xie}}, \bibinfo {author} {\bibfnamefont {L.-M.}\ \bibnamefont {Lu}},\ and\ \bibinfo {author} {\bibfnamefont {Y.-L.}\ \bibnamefont {Tang}},\ }\href {https://doi.org/https://doi.org/10.1016/j.optmat.2021.110952} {\bibfield  {journal} {\bibinfo  {journal} {Opt. Mater.}\ }\textbf {\bibinfo {volume} {114}},\ \bibinfo {pages} {110952} (\bibinfo {year} {2021})}\BibitemShut {NoStop}%
\bibitem [{\citenamefont {Huang}\ \emph {et~al.}(2017)\citenamefont {Huang}, \citenamefont {Jiang},\ and\ \citenamefont {Luo}}]{14}%
  \BibitemOpen
  \bibfield  {author} {\bibinfo {author} {\bibfnamefont {H.-M.}\ \bibnamefont {Huang}}, \bibinfo {author} {\bibfnamefont {Z.-Y.}\ \bibnamefont {Jiang}},\ and\ \bibinfo {author} {\bibfnamefont {S.-J.}\ \bibnamefont {Luo}},\ }\href {https://doi.org/10.1088/1674-1056/26/9/096301} {\bibfield  {journal} {\bibinfo  {journal} {Chin. Phys. B}\ }\textbf {\bibinfo {volume} {26}},\ \bibinfo {pages} {096301} (\bibinfo {year} {2017})}\BibitemShut {NoStop}%
\bibitem [{\citenamefont {Lee}\ \emph {et~al.}(2014)\citenamefont {Lee}, \citenamefont {Stoumpos}, \citenamefont {Zhou}, \citenamefont {Hao}, \citenamefont {Malliakas}, \citenamefont {Yeh}, \citenamefont {Marks}, \citenamefont {Kanatzidis},\ and\ \citenamefont {Chang}}]{16}%
  \BibitemOpen
  \bibfield  {author} {\bibinfo {author} {\bibfnamefont {B.}~\bibnamefont {Lee}}, \bibinfo {author} {\bibfnamefont {C.~C.}\ \bibnamefont {Stoumpos}}, \bibinfo {author} {\bibfnamefont {N.}~\bibnamefont {Zhou}}, \bibinfo {author} {\bibfnamefont {F.}~\bibnamefont {Hao}}, \bibinfo {author} {\bibfnamefont {C.}~\bibnamefont {Malliakas}}, \bibinfo {author} {\bibfnamefont {C.-Y.}\ \bibnamefont {Yeh}}, \bibinfo {author} {\bibfnamefont {T.~J.}\ \bibnamefont {Marks}}, \bibinfo {author} {\bibfnamefont {M.~G.}\ \bibnamefont {Kanatzidis}},\ and\ \bibinfo {author} {\bibfnamefont {R.~P.~H.}\ \bibnamefont {Chang}},\ }\href {https://doi.org/10.1021/ja508464w} {\bibfield  {journal} {\bibinfo  {journal} {J. Am. Chem. Soc.}\ }\textbf {\bibinfo {volume} {136}},\ \bibinfo {pages} {15379} (\bibinfo {year} {2014})},\ \bibinfo {note} {pMID: 25299304},\ \Eprint {https://arxiv.org/abs/https://doi.org/10.1021/ja508464w} {https://doi.org/10.1021/ja508464w} \BibitemShut {NoStop}%
\bibitem [{\citenamefont {Ju}\ \emph {et~al.}(2018)\citenamefont {Ju}, \citenamefont {Chen}, \citenamefont {Zhou}, \citenamefont {Garces}, \citenamefont {Dai}, \citenamefont {Ma}, \citenamefont {Padture},\ and\ \citenamefont {Zeng}}]{17}%
  \BibitemOpen
  \bibfield  {author} {\bibinfo {author} {\bibfnamefont {M.-G.}\ \bibnamefont {Ju}}, \bibinfo {author} {\bibfnamefont {M.}~\bibnamefont {Chen}}, \bibinfo {author} {\bibfnamefont {Y.}~\bibnamefont {Zhou}}, \bibinfo {author} {\bibfnamefont {H.~F.}\ \bibnamefont {Garces}}, \bibinfo {author} {\bibfnamefont {J.}~\bibnamefont {Dai}}, \bibinfo {author} {\bibfnamefont {L.}~\bibnamefont {Ma}}, \bibinfo {author} {\bibfnamefont {N.~P.}\ \bibnamefont {Padture}},\ and\ \bibinfo {author} {\bibfnamefont {X.~C.}\ \bibnamefont {Zeng}},\ }\href {https://doi.org/10.1021/acsenergylett.7b01167} {\bibfield  {journal} {\bibinfo  {journal} {ACS Energy Lett.}\ }\textbf {\bibinfo {volume} {3}},\ \bibinfo {pages} {297} (\bibinfo {year} {2018})},\ \Eprint {https://arxiv.org/abs/https://doi.org/10.1021/acsenergylett.7b01167} {https://doi.org/10.1021/acsenergylett.7b01167} \BibitemShut {NoStop}%
\bibitem [{\citenamefont {Cai}\ \emph {et~al.}(2017)\citenamefont {Cai}, \citenamefont {Xie}, \citenamefont {Ding}, \citenamefont {Chen}, \citenamefont {Thirumal}, \citenamefont {Wong}, \citenamefont {Mathews}, \citenamefont {Mhaisalkar}, \citenamefont {Sherburne},\ and\ \citenamefont {Asta}}]{20}%
  \BibitemOpen
  \bibfield  {author} {\bibinfo {author} {\bibfnamefont {Y.}~\bibnamefont {Cai}}, \bibinfo {author} {\bibfnamefont {W.}~\bibnamefont {Xie}}, \bibinfo {author} {\bibfnamefont {H.}~\bibnamefont {Ding}}, \bibinfo {author} {\bibfnamefont {Y.}~\bibnamefont {Chen}}, \bibinfo {author} {\bibfnamefont {K.}~\bibnamefont {Thirumal}}, \bibinfo {author} {\bibfnamefont {L.~H.}\ \bibnamefont {Wong}}, \bibinfo {author} {\bibfnamefont {N.}~\bibnamefont {Mathews}}, \bibinfo {author} {\bibfnamefont {S.~G.}\ \bibnamefont {Mhaisalkar}}, \bibinfo {author} {\bibfnamefont {M.}~\bibnamefont {Sherburne}},\ and\ \bibinfo {author} {\bibfnamefont {M.}~\bibnamefont {Asta}},\ }\href {https://doi.org/10.1021/acs.chemmater.7b02013} {\bibfield  {journal} {\bibinfo  {journal} {Chem. Mater.}\ }\textbf {\bibinfo {volume} {29}},\ \bibinfo {pages} {7740} (\bibinfo {year} {2017})},\ \Eprint {https://arxiv.org/abs/https://doi.org/10.1021/acs.chemmater.7b02013} {https://doi.org/10.1021/acs.chemmater.7b02013} \BibitemShut {NoStop}%
\bibitem [{\citenamefont {Zhou}\ \emph {et~al.}(2018)\citenamefont {Zhou}, \citenamefont {Liao}, \citenamefont {Huang}, \citenamefont {Wang}, \citenamefont {Xu}, \citenamefont {Chen}, \citenamefont {Kuang},\ and\ \citenamefont {Su}}]{19}%
  \BibitemOpen
  \bibfield  {author} {\bibinfo {author} {\bibfnamefont {L.}~\bibnamefont {Zhou}}, \bibinfo {author} {\bibfnamefont {J.-F.}\ \bibnamefont {Liao}}, \bibinfo {author} {\bibfnamefont {Z.-G.}\ \bibnamefont {Huang}}, \bibinfo {author} {\bibfnamefont {X.-D.}\ \bibnamefont {Wang}}, \bibinfo {author} {\bibfnamefont {Y.-F.}\ \bibnamefont {Xu}}, \bibinfo {author} {\bibfnamefont {H.-Y.}\ \bibnamefont {Chen}}, \bibinfo {author} {\bibfnamefont {D.-B.}\ \bibnamefont {Kuang}},\ and\ \bibinfo {author} {\bibfnamefont {C.-Y.}\ \bibnamefont {Su}},\ }\href {https://doi.org/10.1021/acsenergylett.8b01770} {\bibfield  {journal} {\bibinfo  {journal} {ACS Energy Lett.}\ }\textbf {\bibinfo {volume} {3}},\ \bibinfo {pages} {2613} (\bibinfo {year} {2018})},\ \Eprint {https://arxiv.org/abs/https://doi.org/10.1021/acsenergylett.8b01770} {https://doi.org/10.1021/acsenergylett.8b01770} \BibitemShut {NoStop}%
\bibitem [{\citenamefont {Folgueras}\ \emph {et~al.}(2021)\citenamefont {Folgueras}, \citenamefont {Jin}, \citenamefont {Gao}, \citenamefont {Quan}, \citenamefont {Steele}, \citenamefont {Srivastava}, \citenamefont {Ross}, \citenamefont {Zhang}, \citenamefont {Seeler}, \citenamefont {Schierle-Arndt}, \citenamefont {Asta},\ and\ \citenamefont {Yang}}]{35}%
  \BibitemOpen
  \bibfield  {author} {\bibinfo {author} {\bibfnamefont {M.~C.}\ \bibnamefont {Folgueras}}, \bibinfo {author} {\bibfnamefont {J.}~\bibnamefont {Jin}}, \bibinfo {author} {\bibfnamefont {M.}~\bibnamefont {Gao}}, \bibinfo {author} {\bibfnamefont {L.~N.}\ \bibnamefont {Quan}}, \bibinfo {author} {\bibfnamefont {J.~A.}\ \bibnamefont {Steele}}, \bibinfo {author} {\bibfnamefont {S.}~\bibnamefont {Srivastava}}, \bibinfo {author} {\bibfnamefont {M.~B.}\ \bibnamefont {Ross}}, \bibinfo {author} {\bibfnamefont {R.}~\bibnamefont {Zhang}}, \bibinfo {author} {\bibfnamefont {F.}~\bibnamefont {Seeler}}, \bibinfo {author} {\bibfnamefont {K.}~\bibnamefont {Schierle-Arndt}}, \bibinfo {author} {\bibfnamefont {M.}~\bibnamefont {Asta}},\ and\ \bibinfo {author} {\bibfnamefont {P.}~\bibnamefont {Yang}},\ }\href {https://doi.org/10.1021/acs.jpcc.1c08332} {\bibfield  {journal} {\bibinfo  {journal} {J. Phys. Chem. C}\ }\textbf {\bibinfo {volume} {125}},\ \bibinfo {pages} {25126} (\bibinfo {year} {2021})},\ \Eprint
  {https://arxiv.org/abs/https://doi.org/10.1021/acs.jpcc.1c08332} {https://doi.org/10.1021/acs.jpcc.1c08332} \BibitemShut {NoStop}%
\bibitem [{\citenamefont {Wan}\ \emph {et~al.}(2023)\citenamefont {Wan}, \citenamefont {Jia}, \citenamefont {Dinic}, \citenamefont {Imran}, \citenamefont {Rehl}, \citenamefont {Liu}, \citenamefont {Paritmongkol}, \citenamefont {Xia}, \citenamefont {Wang}, \citenamefont {Liu}, \citenamefont {Wang}, \citenamefont {Lyu}, \citenamefont {Cotella}, \citenamefont {Chun}, \citenamefont {Voznyy}, \citenamefont {Hoogland},\ and\ \citenamefont {Sargent}}]{32}%
  \BibitemOpen
  \bibfield  {author} {\bibinfo {author} {\bibfnamefont {H.}~\bibnamefont {Wan}}, \bibinfo {author} {\bibfnamefont {F.}~\bibnamefont {Jia}}, \bibinfo {author} {\bibfnamefont {F.}~\bibnamefont {Dinic}}, \bibinfo {author} {\bibfnamefont {M.}~\bibnamefont {Imran}}, \bibinfo {author} {\bibfnamefont {B.}~\bibnamefont {Rehl}}, \bibinfo {author} {\bibfnamefont {Y.}~\bibnamefont {Liu}}, \bibinfo {author} {\bibfnamefont {W.}~\bibnamefont {Paritmongkol}}, \bibinfo {author} {\bibfnamefont {P.}~\bibnamefont {Xia}}, \bibinfo {author} {\bibfnamefont {Y.-K.}\ \bibnamefont {Wang}}, \bibinfo {author} {\bibfnamefont {Y.}~\bibnamefont {Liu}}, \bibinfo {author} {\bibfnamefont {S.}~\bibnamefont {Wang}}, \bibinfo {author} {\bibfnamefont {Q.}~\bibnamefont {Lyu}}, \bibinfo {author} {\bibfnamefont {G.~F.}\ \bibnamefont {Cotella}}, \bibinfo {author} {\bibfnamefont {P.}~\bibnamefont {Chun}}, \bibinfo {author} {\bibfnamefont {O.}~\bibnamefont {Voznyy}}, \bibinfo {author} {\bibfnamefont {S.}~\bibnamefont {Hoogland}},\ and\ \bibinfo
  {author} {\bibfnamefont {E.~H.}\ \bibnamefont {Sargent}},\ }\href {https://doi.org/10.1021/acs.chemmater.2c02673} {\bibfield  {journal} {\bibinfo  {journal} {Chem. Mater.}\ }\textbf {\bibinfo {volume} {35}},\ \bibinfo {pages} {948} (\bibinfo {year} {2023})},\ \Eprint {https://arxiv.org/abs/https://doi.org/10.1021/acs.chemmater.2c02673} {https://doi.org/10.1021/acs.chemmater.2c02673} \BibitemShut {NoStop}%
\bibitem [{\citenamefont {Glockzin}\ \emph {et~al.}(2023)\citenamefont {Glockzin}, \citenamefont {Oakley}, \citenamefont {Karmakar}, \citenamefont {Pominov}, \citenamefont {Mitchell}, \citenamefont {Ma}, \citenamefont {Klobukowski},\ and\ \citenamefont {Michaelis}}]{38}%
  \BibitemOpen
  \bibfield  {author} {\bibinfo {author} {\bibfnamefont {B.}~\bibnamefont {Glockzin}}, \bibinfo {author} {\bibfnamefont {M.~S.}\ \bibnamefont {Oakley}}, \bibinfo {author} {\bibfnamefont {A.}~\bibnamefont {Karmakar}}, \bibinfo {author} {\bibfnamefont {A.}~\bibnamefont {Pominov}}, \bibinfo {author} {\bibfnamefont {A.~A.}\ \bibnamefont {Mitchell}}, \bibinfo {author} {\bibfnamefont {X.}~\bibnamefont {Ma}}, \bibinfo {author} {\bibfnamefont {M.}~\bibnamefont {Klobukowski}},\ and\ \bibinfo {author} {\bibfnamefont {V.~K.}\ \bibnamefont {Michaelis}},\ }\href {https://doi.org/10.1021/acs.jpcc.2c08073} {\bibfield  {journal} {\bibinfo  {journal} {J. Phys. Chem. C}\ }\textbf {\bibinfo {volume} {127}},\ \bibinfo {pages} {7284} (\bibinfo {year} {2023})},\ \Eprint {https://arxiv.org/abs/https://doi.org/10.1021/acs.jpcc.2c08073} {https://doi.org/10.1021/acs.jpcc.2c08073} \BibitemShut {NoStop}%
\bibitem [{\citenamefont {Qiu}\ \emph {et~al.}(2017)\citenamefont {Qiu}, \citenamefont {Cao}, \citenamefont {Yuan}, \citenamefont {Chen}, \citenamefont {Qiu}, \citenamefont {Jiang}, \citenamefont {Ye}, \citenamefont {Wang}, \citenamefont {Zeng}, \citenamefont {Liu},\ and\ \citenamefont {Kanatzidis}}]{95}%
  \BibitemOpen
  \bibfield  {author} {\bibinfo {author} {\bibfnamefont {X.}~\bibnamefont {Qiu}}, \bibinfo {author} {\bibfnamefont {B.}~\bibnamefont {Cao}}, \bibinfo {author} {\bibfnamefont {S.}~\bibnamefont {Yuan}}, \bibinfo {author} {\bibfnamefont {X.}~\bibnamefont {Chen}}, \bibinfo {author} {\bibfnamefont {Z.}~\bibnamefont {Qiu}}, \bibinfo {author} {\bibfnamefont {Y.}~\bibnamefont {Jiang}}, \bibinfo {author} {\bibfnamefont {Q.}~\bibnamefont {Ye}}, \bibinfo {author} {\bibfnamefont {H.}~\bibnamefont {Wang}}, \bibinfo {author} {\bibfnamefont {H.}~\bibnamefont {Zeng}}, \bibinfo {author} {\bibfnamefont {J.}~\bibnamefont {Liu}},\ and\ \bibinfo {author} {\bibfnamefont {M.}~\bibnamefont {Kanatzidis}},\ }\href {https://doi.org/10.1016/j.solmat.2016.09.022} {\bibfield  {journal} {\bibinfo  {journal} {Sol. Energy Mater. Sol. Cells}\ }\textbf {\bibinfo {volume} {159}},\ \bibinfo {pages} {227} (\bibinfo {year} {2017})},\ \bibinfo {note} {publisher Copyright: {\textcopyright} 2016 Elsevier B.V.}\BibitemShut {Stop}%
\bibitem [{\citenamefont {Rohlfing}\ and\ \citenamefont {Louie}(1998{\natexlab{a}})}]{45}%
  \BibitemOpen
  \bibfield  {author} {\bibinfo {author} {\bibfnamefont {M.}~\bibnamefont {Rohlfing}}\ and\ \bibinfo {author} {\bibfnamefont {S.~G.}\ \bibnamefont {Louie}},\ }\href {https://doi.org/10.1103/PhysRevLett.81.2312} {\bibfield  {journal} {\bibinfo  {journal} {Phys. Rev. Lett.}\ }\textbf {\bibinfo {volume} {81}},\ \bibinfo {pages} {2312} (\bibinfo {year} {1998}{\natexlab{a}})}\BibitemShut {NoStop}%
\bibitem [{\citenamefont {Heyd}\ \emph {et~al.}(2003)\citenamefont {Heyd}, \citenamefont {Scuseria},\ and\ \citenamefont {Ernzerhof}}]{26}%
  \BibitemOpen
  \bibfield  {author} {\bibinfo {author} {\bibfnamefont {J.}~\bibnamefont {Heyd}}, \bibinfo {author} {\bibfnamefont {G.~E.}\ \bibnamefont {Scuseria}},\ and\ \bibinfo {author} {\bibfnamefont {M.}~\bibnamefont {Ernzerhof}},\ }\href {https://doi.org/10.1063/1.1564060} {\bibfield  {journal} {\bibinfo  {journal} {J. Chem. Phys.}\ }\textbf {\bibinfo {volume} {118}},\ \bibinfo {pages} {8207} (\bibinfo {year} {2003})},\ \Eprint {https://arxiv.org/abs/https://pubs.aip.org/aip/jcp/article-pdf/118/18/8207/10847843/8207\_1\_online.pdf} {https://pubs.aip.org/aip/jcp/article-pdf/118/18/8207/10847843/8207\_1\_online.pdf} \BibitemShut {NoStop}%
\bibitem [{\citenamefont {Hedin}(1965{\natexlab{a}})}]{46}%
  \BibitemOpen
  \bibfield  {author} {\bibinfo {author} {\bibfnamefont {L.}~\bibnamefont {Hedin}},\ }\href {https://doi.org/10.1103/PhysRev.139.A796} {\bibfield  {journal} {\bibinfo  {journal} {Phys. Rev.}\ }\textbf {\bibinfo {volume} {139}},\ \bibinfo {pages} {A796} (\bibinfo {year} {1965}{\natexlab{a}})}\BibitemShut {NoStop}%
\bibitem [{\citenamefont {Kresse}\ and\ \citenamefont {Furthm\"uller}(1996{\natexlab{a}})}]{22}%
  \BibitemOpen
  \bibfield  {author} {\bibinfo {author} {\bibfnamefont {G.}~\bibnamefont {Kresse}}\ and\ \bibinfo {author} {\bibfnamefont {J.}~\bibnamefont {Furthm\"uller}},\ }\href {https://doi.org/10.1103/PhysRevB.54.11169} {\bibfield  {journal} {\bibinfo  {journal} {Phys. Rev. B}\ }\textbf {\bibinfo {volume} {54}},\ \bibinfo {pages} {11169} (\bibinfo {year} {1996}{\natexlab{a}})}\BibitemShut {NoStop}%
\bibitem [{\citenamefont {Kresse}\ and\ \citenamefont {Furthm\"uller}(1996{\natexlab{b}})}]{23}%
  \BibitemOpen
  \bibfield  {author} {\bibinfo {author} {\bibfnamefont {G.}~\bibnamefont {Kresse}}\ and\ \bibinfo {author} {\bibfnamefont {J.}~\bibnamefont {Furthm\"uller}},\ }\href {https://doi.org/https://doi.org/10.1016/0927-0256(96)00008-0} {\bibfield  {journal} {\bibinfo  {journal} {Comput. Mater. Sci.}\ }\textbf {\bibinfo {volume} {6}},\ \bibinfo {pages} {15} (\bibinfo {year} {1996}{\natexlab{b}})}\BibitemShut {NoStop}%
\bibitem [{\citenamefont {Mortensen}\ \emph {et~al.}(2005)\citenamefont {Mortensen}, \citenamefont {Hansen},\ and\ \citenamefont {Jacobsen}}]{24}%
  \BibitemOpen
  \bibfield  {author} {\bibinfo {author} {\bibfnamefont {J.~J.}\ \bibnamefont {Mortensen}}, \bibinfo {author} {\bibfnamefont {L.~B.}\ \bibnamefont {Hansen}},\ and\ \bibinfo {author} {\bibfnamefont {K.~W.}\ \bibnamefont {Jacobsen}},\ }\href {https://doi.org/10.1103/PhysRevB.71.035109} {\bibfield  {journal} {\bibinfo  {journal} {Phys. Rev. B}\ }\textbf {\bibinfo {volume} {71}},\ \bibinfo {pages} {035109} (\bibinfo {year} {2005})}\BibitemShut {NoStop}%
\bibitem [{\citenamefont {Perdew}\ \emph {et~al.}(1996)\citenamefont {Perdew}, \citenamefont {Burke},\ and\ \citenamefont {Ernzerhof}}]{25}%
  \BibitemOpen
  \bibfield  {author} {\bibinfo {author} {\bibfnamefont {J.~P.}\ \bibnamefont {Perdew}}, \bibinfo {author} {\bibfnamefont {K.}~\bibnamefont {Burke}},\ and\ \bibinfo {author} {\bibfnamefont {M.}~\bibnamefont {Ernzerhof}},\ }\href {https://doi.org/10.1103/PhysRevLett.77.3865} {\bibfield  {journal} {\bibinfo  {journal} {Phys. Rev. Lett.}\ }\textbf {\bibinfo {volume} {77}},\ \bibinfo {pages} {3865} (\bibinfo {year} {1996})}\BibitemShut {NoStop}%
\bibitem [{\citenamefont {Hedin}(1965{\natexlab{b}})}]{29}%
  \BibitemOpen
  \bibfield  {author} {\bibinfo {author} {\bibfnamefont {L.}~\bibnamefont {Hedin}},\ }\href {https://doi.org/10.1103/PhysRev.139.A796} {\bibfield  {journal} {\bibinfo  {journal} {Phys. Rev.}\ }\textbf {\bibinfo {volume} {139}},\ \bibinfo {pages} {A796} (\bibinfo {year} {1965}{\natexlab{b}})}\BibitemShut {NoStop}%
\bibitem [{\citenamefont {Hybertsen}\ and\ \citenamefont {Louie}(1985)}]{30}%
  \BibitemOpen
  \bibfield  {author} {\bibinfo {author} {\bibfnamefont {M.~S.}\ \bibnamefont {Hybertsen}}\ and\ \bibinfo {author} {\bibfnamefont {S.~G.}\ \bibnamefont {Louie}},\ }\href {https://doi.org/10.1103/PhysRevLett.55.1418} {\bibfield  {journal} {\bibinfo  {journal} {Phys. Rev. Lett.}\ }\textbf {\bibinfo {volume} {55}},\ \bibinfo {pages} {1418} (\bibinfo {year} {1985})}\BibitemShut {NoStop}%
\bibitem [{\citenamefont {Albrecht}\ \emph {et~al.}(1998)\citenamefont {Albrecht}, \citenamefont {Reining}, \citenamefont {Del~Sole},\ and\ \citenamefont {Onida}}]{27}%
  \BibitemOpen
  \bibfield  {author} {\bibinfo {author} {\bibfnamefont {S.}~\bibnamefont {Albrecht}}, \bibinfo {author} {\bibfnamefont {L.}~\bibnamefont {Reining}}, \bibinfo {author} {\bibfnamefont {R.}~\bibnamefont {Del~Sole}},\ and\ \bibinfo {author} {\bibfnamefont {G.}~\bibnamefont {Onida}},\ }\href {https://doi.org/10.1103/PhysRevLett.80.4510} {\bibfield  {journal} {\bibinfo  {journal} {Phys. Rev. Lett.}\ }\textbf {\bibinfo {volume} {80}},\ \bibinfo {pages} {4510} (\bibinfo {year} {1998})}\BibitemShut {NoStop}%
\bibitem [{\citenamefont {Rohlfing}\ and\ \citenamefont {Louie}(1998{\natexlab{b}})}]{28}%
  \BibitemOpen
  \bibfield  {author} {\bibinfo {author} {\bibfnamefont {M.}~\bibnamefont {Rohlfing}}\ and\ \bibinfo {author} {\bibfnamefont {S.~G.}\ \bibnamefont {Louie}},\ }\href {https://doi.org/10.1103/PhysRevLett.81.2312} {\bibfield  {journal} {\bibinfo  {journal} {Phys. Rev. Lett.}\ }\textbf {\bibinfo {volume} {81}},\ \bibinfo {pages} {2312} (\bibinfo {year} {1998}{\natexlab{b}})}\BibitemShut {NoStop}%
\bibitem [{\citenamefont {Gajdo\ifmmode~\check{s}\else \v{s}\fi{}}\ \emph {et~al.}(2006)\citenamefont {Gajdo\ifmmode~\check{s}\else \v{s}\fi{}}, \citenamefont {Hummer}, \citenamefont {Kresse}, \citenamefont {Furthm\"uller},\ and\ \citenamefont {Bechstedt}}]{31}%
  \BibitemOpen
  \bibfield  {author} {\bibinfo {author} {\bibfnamefont {M.}~\bibnamefont {Gajdo\ifmmode~\check{s}\else \v{s}\fi{}}}, \bibinfo {author} {\bibfnamefont {K.}~\bibnamefont {Hummer}}, \bibinfo {author} {\bibfnamefont {G.}~\bibnamefont {Kresse}}, \bibinfo {author} {\bibfnamefont {J.}~\bibnamefont {Furthm\"uller}},\ and\ \bibinfo {author} {\bibfnamefont {F.}~\bibnamefont {Bechstedt}},\ }\href {https://doi.org/10.1103/PhysRevB.73.045112} {\bibfield  {journal} {\bibinfo  {journal} {Phys. Rev. B}\ }\textbf {\bibinfo {volume} {73}},\ \bibinfo {pages} {045112} (\bibinfo {year} {2006})}\BibitemShut {NoStop}%
\bibitem [{\citenamefont {Thiele}\ \emph {et~al.}(1983)\citenamefont {Thiele}, \citenamefont {Mrozek}, \citenamefont {K\"ammerer},\ and\ \citenamefont {Wittmann}}]{59}%
  \BibitemOpen
  \bibfield  {author} {\bibinfo {author} {\bibfnamefont {G.}~\bibnamefont {Thiele}}, \bibinfo {author} {\bibfnamefont {C.}~\bibnamefont {Mrozek}}, \bibinfo {author} {\bibfnamefont {D.}~\bibnamefont {K\"ammerer}},\ and\ \bibinfo {author} {\bibfnamefont {K.}~\bibnamefont {Wittmann}},\ }\href {https://doi.org/doi:10.1515/znb-1983-0802} {\bibfield  {journal} {\bibinfo  {journal} {, Z. Naturforsch. 38b}\ }\textbf {\bibinfo {volume} {38}},\ \bibinfo {pages} {905} (\bibinfo {year} {1983})}\BibitemShut {NoStop}%
\bibitem [{\citenamefont {Torres}\ \emph {et~al.}(1997)\citenamefont {Torres}, \citenamefont {Freire},\ and\ \citenamefont {Katiyar}}]{60}%
  \BibitemOpen
  \bibfield  {author} {\bibinfo {author} {\bibfnamefont {D.~I.}\ \bibnamefont {Torres}}, \bibinfo {author} {\bibfnamefont {J.~D.}\ \bibnamefont {Freire}},\ and\ \bibinfo {author} {\bibfnamefont {R.~S.}\ \bibnamefont {Katiyar}},\ }\href {https://doi.org/10.1103/PhysRevB.56.7763} {\bibfield  {journal} {\bibinfo  {journal} {Phys. Rev. B}\ }\textbf {\bibinfo {volume} {56}},\ \bibinfo {pages} {7763} (\bibinfo {year} {1997})}\BibitemShut {NoStop}%
\bibitem [{\citenamefont {Ketelaar}\ \emph {et~al.}(1937)\citenamefont {Ketelaar}, \citenamefont {Rietdijk},\ and\ \citenamefont {Van~Staveren}}]{62}%
  \BibitemOpen
  \bibfield  {author} {\bibinfo {author} {\bibfnamefont {J.}~\bibnamefont {Ketelaar}}, \bibinfo {author} {\bibfnamefont {A.}~\bibnamefont {Rietdijk}},\ and\ \bibinfo {author} {\bibfnamefont {C.}~\bibnamefont {Van~Staveren}},\ }\href@noop {} {\bibfield  {journal} {\bibinfo  {journal} {Recl. Trav. Chim. Pays-Bas}\ }\textbf {\bibinfo {volume} {56}},\ \bibinfo {pages} {907} (\bibinfo {year} {1937})}\BibitemShut {NoStop}%
\bibitem [{\citenamefont {Werker}(1939)}]{63}%
  \BibitemOpen
  \bibfield  {author} {\bibinfo {author} {\bibfnamefont {W.}~\bibnamefont {Werker}},\ }\href@noop {} {\bibfield  {journal} {\bibinfo  {journal} {Recl. Trav. Chim. Pays-Bas}\ }\textbf {\bibinfo {volume} {58}},\ \bibinfo {pages} {257} (\bibinfo {year} {1939})}\BibitemShut {NoStop}%
\bibitem [{\citenamefont {Engel}(1935)}]{84}%
  \BibitemOpen
  \bibfield  {author} {\bibinfo {author} {\bibfnamefont {G.}~\bibnamefont {Engel}},\ }\href {https://doi.org/doi:10.1524/zkri.1935.90.1.341} {\bibfield  {journal} {\bibinfo  {journal} {Z. Kristallogr Cryst. Mater.}\ }\textbf {\bibinfo {volume} {90}},\ \bibinfo {pages} {341} (\bibinfo {year} {1935})}\BibitemShut {NoStop}%
\bibitem [{\citenamefont {Mahmood}\ \emph {et~al.}(2021)\citenamefont {Mahmood}, \citenamefont {Hassan}, \citenamefont {Flemban}, \citenamefont {{Ul Haq}}, \citenamefont {AlFaify}, \citenamefont {Kattan},\ and\ \citenamefont {Laref}}]{83}%
  \BibitemOpen
  \bibfield  {author} {\bibinfo {author} {\bibfnamefont {Q.}~\bibnamefont {Mahmood}}, \bibinfo {author} {\bibfnamefont {M.}~\bibnamefont {Hassan}}, \bibinfo {author} {\bibfnamefont {T.~H.}\ \bibnamefont {Flemban}}, \bibinfo {author} {\bibfnamefont {B.}~\bibnamefont {{Ul Haq}}}, \bibinfo {author} {\bibfnamefont {S.}~\bibnamefont {AlFaify}}, \bibinfo {author} {\bibfnamefont {N.~A.}\ \bibnamefont {Kattan}},\ and\ \bibinfo {author} {\bibfnamefont {A.}~\bibnamefont {Laref}},\ }\href {https://doi.org/https://doi.org/10.1016/j.jpcs.2020.109665} {\bibfield  {journal} {\bibinfo  {journal} {J. Phys. Chem. Solids}\ }\textbf {\bibinfo {volume} {148}},\ \bibinfo {pages} {109665} (\bibinfo {year} {2021})}\BibitemShut {NoStop}%
\bibitem [{\citenamefont {Suzuki}\ and\ \citenamefont {Tsuyama}(2021)}]{85}%
  \BibitemOpen
  \bibfield  {author} {\bibinfo {author} {\bibfnamefont {S.}~\bibnamefont {Suzuki}}\ and\ \bibinfo {author} {\bibfnamefont {M.}~\bibnamefont {Tsuyama}},\ }\href {https://doi.org/https://doi.org/10.1016/j.optmat.2021.111323} {\bibfield  {journal} {\bibinfo  {journal} {Opt. Mater.}\ }\textbf {\bibinfo {volume} {119}},\ \bibinfo {pages} {111323} (\bibinfo {year} {2021})}\BibitemShut {NoStop}%
\bibitem [{\citenamefont {de~Gironcoli}(1995)}]{64}%
  \BibitemOpen
  \bibfield  {author} {\bibinfo {author} {\bibfnamefont {S.}~\bibnamefont {de~Gironcoli}},\ }\href {https://doi.org/10.1103/PhysRevB.51.6773} {\bibfield  {journal} {\bibinfo  {journal} {Phys. Rev. B}\ }\textbf {\bibinfo {volume} {51}},\ \bibinfo {pages} {6773} (\bibinfo {year} {1995})}\BibitemShut {NoStop}%
\bibitem [{\citenamefont {Mouhat}\ and\ \citenamefont {Coudert}(2014)}]{51}%
  \BibitemOpen
  \bibfield  {author} {\bibinfo {author} {\bibfnamefont {F.}~\bibnamefont {Mouhat}}\ and\ \bibinfo {author} {\bibfnamefont {F.~m. c.-X.}\ \bibnamefont {Coudert}},\ }\href {https://doi.org/10.1103/PhysRevB.90.224104} {\bibfield  {journal} {\bibinfo  {journal} {Phys. Rev. B}\ }\textbf {\bibinfo {volume} {90}},\ \bibinfo {pages} {224104} (\bibinfo {year} {2014})}\BibitemShut {NoStop}%
\bibitem [{\citenamefont {Wang}\ \emph {et~al.}(2021)\citenamefont {Wang}, \citenamefont {Xu}, \citenamefont {Liu}, \citenamefont {Tang},\ and\ \citenamefont {Geng}}]{52}%
  \BibitemOpen
  \bibfield  {author} {\bibinfo {author} {\bibfnamefont {V.}~\bibnamefont {Wang}}, \bibinfo {author} {\bibfnamefont {N.}~\bibnamefont {Xu}}, \bibinfo {author} {\bibfnamefont {J.-C.}\ \bibnamefont {Liu}}, \bibinfo {author} {\bibfnamefont {G.}~\bibnamefont {Tang}},\ and\ \bibinfo {author} {\bibfnamefont {W.-T.}\ \bibnamefont {Geng}},\ }\href {https://doi.org/https://doi.org/10.1016/j.cpc.2021.108033} {\bibfield  {journal} {\bibinfo  {journal} {Comput. Phys. Commun.}\ }\textbf {\bibinfo {volume} {267}},\ \bibinfo {pages} {108033} (\bibinfo {year} {2021})}\BibitemShut {NoStop}%
\bibitem [{\citenamefont {Wu}\ \emph {et~al.}(2007)\citenamefont {Wu}, \citenamefont {Zhao}, \citenamefont {Xiang}, \citenamefont {Hao}, \citenamefont {Liu},\ and\ \citenamefont {Meng}}]{53}%
  \BibitemOpen
  \bibfield  {author} {\bibinfo {author} {\bibfnamefont {Z.-j.}\ \bibnamefont {Wu}}, \bibinfo {author} {\bibfnamefont {E.-j.}\ \bibnamefont {Zhao}}, \bibinfo {author} {\bibfnamefont {H.-p.}\ \bibnamefont {Xiang}}, \bibinfo {author} {\bibfnamefont {X.-f.}\ \bibnamefont {Hao}}, \bibinfo {author} {\bibfnamefont {X.-j.}\ \bibnamefont {Liu}},\ and\ \bibinfo {author} {\bibfnamefont {J.}~\bibnamefont {Meng}},\ }\href {https://doi.org/10.1103/PhysRevB.76.054115} {\bibfield  {journal} {\bibinfo  {journal} {Phys. Rev. B}\ }\textbf {\bibinfo {volume} {76}},\ \bibinfo {pages} {054115} (\bibinfo {year} {2007})}\BibitemShut {NoStop}%
\bibitem [{\citenamefont {Hill}(1952)}]{54}%
  \BibitemOpen
  \bibfield  {author} {\bibinfo {author} {\bibfnamefont {R.}~\bibnamefont {Hill}},\ }\href {https://doi.org/10.1088/0370-1298/65/5/307} {\bibfield  {journal} {\bibinfo  {journal} {Proc. Phys. Soc. A}\ }\textbf {\bibinfo {volume} {65}},\ \bibinfo {pages} {349} (\bibinfo {year} {1952})}\BibitemShut {NoStop}%
\bibitem [{\citenamefont {Pugh}(1954)}]{55}%
  \BibitemOpen
  \bibfield  {author} {\bibinfo {author} {\bibfnamefont {S.}~\bibnamefont {Pugh}},\ }\href {https://doi.org/10.1080/14786440808520496} {\bibfield  {journal} {\bibinfo  {journal} {S. Pugh}\ }\textbf {\bibinfo {volume} {45}},\ \bibinfo {pages} {823} (\bibinfo {year} {1954})},\ \Eprint {https://arxiv.org/abs/https://doi.org/10.1080/14786440808520496} {https://doi.org/10.1080/14786440808520496} \BibitemShut {NoStop}%
\bibitem [{\citenamefont {Huang}\ \emph {et~al.}(2015)\citenamefont {Huang}, \citenamefont {Duan}, \citenamefont {Hu}, \citenamefont {Sun},\ and\ \citenamefont {Chen}}]{56}%
  \BibitemOpen
  \bibfield  {author} {\bibinfo {author} {\bibfnamefont {B.}~\bibnamefont {Huang}}, \bibinfo {author} {\bibfnamefont {Y.-H.}\ \bibnamefont {Duan}}, \bibinfo {author} {\bibfnamefont {W.-C.}\ \bibnamefont {Hu}}, \bibinfo {author} {\bibfnamefont {Y.}~\bibnamefont {Sun}},\ and\ \bibinfo {author} {\bibfnamefont {S.}~\bibnamefont {Chen}},\ }\href {https://doi.org/https://doi.org/10.1016/j.ceramint.2015.01.132} {\bibfield  {journal} {\bibinfo  {journal} {Ceram. Int.}\ }\textbf {\bibinfo {volume} {41}},\ \bibinfo {pages} {6831} (\bibinfo {year} {2015})}\BibitemShut {NoStop}%
\bibitem [{\citenamefont {Zhang}\ \emph {et~al.}(2007)\citenamefont {Zhang}, \citenamefont {Zhang}, \citenamefont {Xu},\ and\ \citenamefont {Ji}}]{57}%
  \BibitemOpen
  \bibfield  {author} {\bibinfo {author} {\bibfnamefont {J.-M.}\ \bibnamefont {Zhang}}, \bibinfo {author} {\bibfnamefont {Y.}~\bibnamefont {Zhang}}, \bibinfo {author} {\bibfnamefont {K.-W.}\ \bibnamefont {Xu}},\ and\ \bibinfo {author} {\bibfnamefont {V.}~\bibnamefont {Ji}},\ }\href {https://doi.org/https://doi.org/10.1016/j.jpcs.2007.01.025} {\bibfield  {journal} {\bibinfo  {journal} {J. Phys. Chem. Solids}\ }\textbf {\bibinfo {volume} {68}},\ \bibinfo {pages} {503} (\bibinfo {year} {2007})}\BibitemShut {NoStop}%
\bibitem [{\citenamefont {Ranganathan}\ and\ \citenamefont {Ostoja-Starzewski}(2008)}]{58}%
  \BibitemOpen
  \bibfield  {author} {\bibinfo {author} {\bibfnamefont {S.~I.}\ \bibnamefont {Ranganathan}}\ and\ \bibinfo {author} {\bibfnamefont {M.}~\bibnamefont {Ostoja-Starzewski}},\ }\href {https://doi.org/10.1103/PhysRevLett.101.055504} {\bibfield  {journal} {\bibinfo  {journal} {Phys. Rev. Lett.}\ }\textbf {\bibinfo {volume} {101}},\ \bibinfo {pages} {055504} (\bibinfo {year} {2008})}\BibitemShut {NoStop}%
\bibitem [{\citenamefont {Ju}\ \emph {et~al.}(2017)\citenamefont {Ju}, \citenamefont {Dai}, \citenamefont {Ma},\ and\ \citenamefont {Zeng}}]{69}%
  \BibitemOpen
  \bibfield  {author} {\bibinfo {author} {\bibfnamefont {M.-G.}\ \bibnamefont {Ju}}, \bibinfo {author} {\bibfnamefont {J.}~\bibnamefont {Dai}}, \bibinfo {author} {\bibfnamefont {L.}~\bibnamefont {Ma}},\ and\ \bibinfo {author} {\bibfnamefont {X.~C.}\ \bibnamefont {Zeng}},\ }\href {https://doi.org/https://doi.org/10.1002/aenm.201700216} {\bibfield  {journal} {\bibinfo  {journal} {Adv. Energy Mater.}\ }\textbf {\bibinfo {volume} {7}},\ \bibinfo {pages} {1700216} (\bibinfo {year} {2017})},\ \Eprint {https://arxiv.org/abs/https://onlinelibrary.wiley.com/doi/pdf/10.1002/aenm.201700216} {https://onlinelibrary.wiley.com/doi/pdf/10.1002/aenm.201700216} \BibitemShut {NoStop}%
\bibitem [{\citenamefont {Li}\ \emph {et~al.}(2019)\citenamefont {Li}, \citenamefont {Zhou}, \citenamefont {Ge}, \citenamefont {Ren}, \citenamefont {Zhao}, \citenamefont {Wan}, \citenamefont {Zhang},\ and\ \citenamefont {Liu}}]{87}%
  \BibitemOpen
  \bibfield  {author} {\bibinfo {author} {\bibfnamefont {Q.}~\bibnamefont {Li}}, \bibinfo {author} {\bibfnamefont {L.}~\bibnamefont {Zhou}}, \bibinfo {author} {\bibfnamefont {Y.}~\bibnamefont {Ge}}, \bibinfo {author} {\bibfnamefont {Y.}~\bibnamefont {Ren}}, \bibinfo {author} {\bibfnamefont {J.}~\bibnamefont {Zhao}}, \bibinfo {author} {\bibfnamefont {W.}~\bibnamefont {Wan}}, \bibinfo {author} {\bibfnamefont {K.}~\bibnamefont {Zhang}},\ and\ \bibinfo {author} {\bibfnamefont {Y.}~\bibnamefont {Liu}},\ }\href {https://arxiv.org/abs/1908.02187} {\bibfield  {journal} {\bibinfo  {journal} {arXiv preprint arXiv:1908.02187}\ } (\bibinfo {year} {2019})},\ \Eprint {https://arxiv.org/abs/1908.02187} {arXiv:1908.02187 [physics.app-ph]} \BibitemShut {NoStop}%
\bibitem [{\citenamefont {Al-Muhimeed}\ \emph {et~al.}(2021)\citenamefont {Al-Muhimeed}, \citenamefont {Shafique}, \citenamefont {AlObaid}, \citenamefont {Morsi}, \citenamefont {Nazir}, \citenamefont {AL-Anazy},\ and\ \citenamefont {Mahmood}}]{88}%
  \BibitemOpen
  \bibfield  {author} {\bibinfo {author} {\bibfnamefont {T.~I.}\ \bibnamefont {Al-Muhimeed}}, \bibinfo {author} {\bibfnamefont {A.}~\bibnamefont {Shafique}}, \bibinfo {author} {\bibfnamefont {A.~A.}\ \bibnamefont {AlObaid}}, \bibinfo {author} {\bibfnamefont {M.}~\bibnamefont {Morsi}}, \bibinfo {author} {\bibfnamefont {G.}~\bibnamefont {Nazir}}, \bibinfo {author} {\bibfnamefont {M.~m.}\ \bibnamefont {AL-Anazy}},\ and\ \bibinfo {author} {\bibfnamefont {Q.}~\bibnamefont {Mahmood}},\ }\href {https://doi.org/https://doi.org/10.1002/er.7022} {\bibfield  {journal} {\bibinfo  {journal} {Int. J. Energy Res.}\ }\textbf {\bibinfo {volume} {45}},\ \bibinfo {pages} {19645} (\bibinfo {year} {2021})},\ \Eprint {https://arxiv.org/abs/https://onlinelibrary.wiley.com/doi/pdf/10.1002/er.7022} {https://onlinelibrary.wiley.com/doi/pdf/10.1002/er.7022} \BibitemShut {NoStop}%
\bibitem [{\citenamefont {Chadli}\ \emph {et~al.}(2022)\citenamefont {Chadli}, \citenamefont {{Bekhti Siad}}, \citenamefont {Baira}, \citenamefont {Siad}, \citenamefont {Allouche},\ and\ \citenamefont {Reguig}}]{90}%
  \BibitemOpen
  \bibfield  {author} {\bibinfo {author} {\bibfnamefont {S.}~\bibnamefont {Chadli}}, \bibinfo {author} {\bibfnamefont {A.}~\bibnamefont {{Bekhti Siad}}}, \bibinfo {author} {\bibfnamefont {M.}~\bibnamefont {Baira}}, \bibinfo {author} {\bibfnamefont {M.}~\bibnamefont {Siad}}, \bibinfo {author} {\bibfnamefont {A.}~\bibnamefont {Allouche}},\ and\ \bibinfo {author} {\bibfnamefont {A.}~\bibnamefont {Reguig}},\ }\href {https://doi.org/https://doi.org/10.1016/j.ssc.2021.114633} {\bibfield  {journal} {\bibinfo  {journal} {Solid State Commun.}\ }\textbf {\bibinfo {volume} {342}},\ \bibinfo {pages} {114633} (\bibinfo {year} {2022})}\BibitemShut {NoStop}%
\bibitem [{\citenamefont {Qamar}\ \emph {et~al.}(2022)\citenamefont {Qamar}, \citenamefont {Lin}, \citenamefont {Tsai},\ and\ \citenamefont {Lin}}]{89}%
  \BibitemOpen
  \bibfield  {author} {\bibinfo {author} {\bibfnamefont {S.~A.}\ \bibnamefont {Qamar}}, \bibinfo {author} {\bibfnamefont {T.-W.}\ \bibnamefont {Lin}}, \bibinfo {author} {\bibfnamefont {Y.-T.}\ \bibnamefont {Tsai}},\ and\ \bibinfo {author} {\bibfnamefont {C.~C.}\ \bibnamefont {Lin}},\ }\href {https://doi.org/10.1021/acsanm.2c01647} {\bibfield  {journal} {\bibinfo  {journal} {ACS Appl. Nano Mater.}\ }\textbf {\bibinfo {volume} {5}},\ \bibinfo {pages} {7580} (\bibinfo {year} {2022})},\ \Eprint {https://arxiv.org/abs/https://doi.org/10.1021/acsanm.2c01647} {https://doi.org/10.1021/acsanm.2c01647} \BibitemShut {NoStop}%
\bibitem [{\citenamefont {Maughan}\ \emph {et~al.}(2018)\citenamefont {Maughan}, \citenamefont {Ganose}, \citenamefont {Almaker}, \citenamefont {Scanlon},\ and\ \citenamefont {Neilson}}]{65}%
  \BibitemOpen
  \bibfield  {author} {\bibinfo {author} {\bibfnamefont {A.~E.}\ \bibnamefont {Maughan}}, \bibinfo {author} {\bibfnamefont {A.~M.}\ \bibnamefont {Ganose}}, \bibinfo {author} {\bibfnamefont {M.~A.}\ \bibnamefont {Almaker}}, \bibinfo {author} {\bibfnamefont {D.~O.}\ \bibnamefont {Scanlon}},\ and\ \bibinfo {author} {\bibfnamefont {J.~R.}\ \bibnamefont {Neilson}},\ }\href {https://doi.org/10.1021/acs.chemmater.8b01549} {\bibfield  {journal} {\bibinfo  {journal} {Chem. Mater.}\ }\textbf {\bibinfo {volume} {30}},\ \bibinfo {pages} {3909} (\bibinfo {year} {2018})},\ \Eprint {https://arxiv.org/abs/https://doi.org/10.1021/acs.chemmater.8b01549} {https://doi.org/10.1021/acs.chemmater.8b01549} \BibitemShut {NoStop}%
\bibitem [{\citenamefont {Hemidi}\ \emph {et~al.}(2023)\citenamefont {Hemidi}, \citenamefont {Seddik}, \citenamefont {Benmessabih}, \citenamefont {Batouche}, \citenamefont {Ouerghui}, \citenamefont {Abdallah}, \citenamefont {Surucu},\ and\ \citenamefont {Ahmad}}]{91}%
  \BibitemOpen
  \bibfield  {author} {\bibinfo {author} {\bibfnamefont {D.}~\bibnamefont {Hemidi}}, \bibinfo {author} {\bibfnamefont {T.}~\bibnamefont {Seddik}}, \bibinfo {author} {\bibfnamefont {T.}~\bibnamefont {Benmessabih}}, \bibinfo {author} {\bibfnamefont {M.}~\bibnamefont {Batouche}}, \bibinfo {author} {\bibfnamefont {W.}~\bibnamefont {Ouerghui}}, \bibinfo {author} {\bibfnamefont {H.~B.}\ \bibnamefont {Abdallah}}, \bibinfo {author} {\bibfnamefont {G.}~\bibnamefont {Surucu}},\ and\ \bibinfo {author} {\bibfnamefont {S.}~\bibnamefont {Ahmad}},\ }\href {https://doi.org/10.1007/s00339-023-07034-w} {\bibfield  {journal} {\bibinfo  {journal} {Appl. Phys. A: Mater. Sci. Process.}\ }\textbf {\bibinfo {volume} {129}},\ \bibinfo {pages} {762} (\bibinfo {year} {2023})}\BibitemShut {NoStop}%
\bibitem [{\citenamefont {Kumar}\ \emph {et~al.}(2021)\citenamefont {Kumar}, \citenamefont {Singh}, \citenamefont {Gill},\ and\ \citenamefont {Bhattacharya}}]{74}%
  \BibitemOpen
  \bibfield  {author} {\bibinfo {author} {\bibfnamefont {M.}~\bibnamefont {Kumar}}, \bibinfo {author} {\bibfnamefont {A.}~\bibnamefont {Singh}}, \bibinfo {author} {\bibfnamefont {D.}~\bibnamefont {Gill}},\ and\ \bibinfo {author} {\bibfnamefont {S.}~\bibnamefont {Bhattacharya}},\ }\href {https://doi.org/10.1021/acs.jpclett.1c01034} {\bibfield  {journal} {\bibinfo  {journal} {J. Phys. Chem. Lett.}\ }\textbf {\bibinfo {volume} {12}},\ \bibinfo {pages} {5301} (\bibinfo {year} {2021})},\ \bibinfo {note} {pMID: 34061540},\ \Eprint {https://arxiv.org/abs/https://doi.org/10.1021/acs.jpclett.1c01034} {https://doi.org/10.1021/acs.jpclett.1c01034} \BibitemShut {NoStop}%
\bibitem [{\citenamefont {Bokdam}\ \emph {et~al.}(2016)\citenamefont {Bokdam}, \citenamefont {Sander}, \citenamefont {Stroppa}, \citenamefont {Picozzi}, \citenamefont {Sarma}, \citenamefont {Franchini},\ and\ \citenamefont {Kresse}}]{79}%
  \BibitemOpen
  \bibfield  {author} {\bibinfo {author} {\bibfnamefont {M.}~\bibnamefont {Bokdam}}, \bibinfo {author} {\bibfnamefont {T.}~\bibnamefont {Sander}}, \bibinfo {author} {\bibfnamefont {A.}~\bibnamefont {Stroppa}}, \bibinfo {author} {\bibfnamefont {S.}~\bibnamefont {Picozzi}}, \bibinfo {author} {\bibfnamefont {D.~D.}\ \bibnamefont {Sarma}}, \bibinfo {author} {\bibfnamefont {C.}~\bibnamefont {Franchini}},\ and\ \bibinfo {author} {\bibfnamefont {G.}~\bibnamefont {Kresse}},\ }\href {https://doi.org/10.1038/srep28618} {\bibfield  {journal} {\bibinfo  {journal} {Sci. Rep.}\ }\textbf {\bibinfo {volume} {6}},\ \bibinfo {pages} {28618} (\bibinfo {year} {2016})}\BibitemShut {NoStop}%
\bibitem [{\citenamefont {Wang}\ \emph {et~al.}(2017)\citenamefont {Wang}, \citenamefont {Meng},\ and\ \citenamefont {Yan}}]{75}%
  \BibitemOpen
  \bibfield  {author} {\bibinfo {author} {\bibfnamefont {X.}~\bibnamefont {Wang}}, \bibinfo {author} {\bibfnamefont {W.}~\bibnamefont {Meng}},\ and\ \bibinfo {author} {\bibfnamefont {Y.}~\bibnamefont {Yan}},\ }\bibfield  {journal} {\bibinfo  {journal} {J. Appl. Phys.}\ }\textbf {\bibinfo {volume} {122}},\ \href {https://doi.org/10.1063/1.4991913} {10.1063/1.4991913} (\bibinfo {year} {2017}),\ \Eprint {https://arxiv.org/abs/https://pubs.aip.org/aip/jap/article-pdf/doi/10.1063/1.4991913/15201223/085104\_1\_online.pdf} {https://pubs.aip.org/aip/jap/article-pdf/doi/10.1063/1.4991913/15201223/085104\_1\_online.pdf} \BibitemShut {NoStop}%
\bibitem [{\citenamefont {Ferreira}\ \emph {et~al.}(2019)\citenamefont {Ferreira}, \citenamefont {Chaves}, \citenamefont {Peres},\ and\ \citenamefont {Ribeiro}}]{76}%
  \BibitemOpen
  \bibfield  {author} {\bibinfo {author} {\bibfnamefont {F.}~\bibnamefont {Ferreira}}, \bibinfo {author} {\bibfnamefont {A.~J.}\ \bibnamefont {Chaves}}, \bibinfo {author} {\bibfnamefont {N.~M.~R.}\ \bibnamefont {Peres}},\ and\ \bibinfo {author} {\bibfnamefont {R.~M.}\ \bibnamefont {Ribeiro}},\ }\href {https://doi.org/10.1364/JOSAB.36.000674} {\bibfield  {journal} {\bibinfo  {journal} {J. Opt. Soc. Am. B}\ }\textbf {\bibinfo {volume} {36}},\ \bibinfo {pages} {674} (\bibinfo {year} {2019})}\BibitemShut {NoStop}%
\bibitem [{\citenamefont {Filip}\ \emph {et~al.}(2021)\citenamefont {Filip}, \citenamefont {Haber},\ and\ \citenamefont {Neaton}}]{82}%
  \BibitemOpen
  \bibfield  {author} {\bibinfo {author} {\bibfnamefont {M.~R.}\ \bibnamefont {Filip}}, \bibinfo {author} {\bibfnamefont {J.~B.}\ \bibnamefont {Haber}},\ and\ \bibinfo {author} {\bibfnamefont {J.~B.}\ \bibnamefont {Neaton}},\ }\href {https://doi.org/10.1103/PhysRevLett.127.067401} {\bibfield  {journal} {\bibinfo  {journal} {Phys. Rev. Lett.}\ }\textbf {\bibinfo {volume} {127}},\ \bibinfo {pages} {067401} (\bibinfo {year} {2021})}\BibitemShut {NoStop}%
\bibitem [{\citenamefont {Hellwarth}\ and\ \citenamefont {Biaggio}(1999)}]{78}%
  \BibitemOpen
  \bibfield  {author} {\bibinfo {author} {\bibfnamefont {R.~W.}\ \bibnamefont {Hellwarth}}\ and\ \bibinfo {author} {\bibfnamefont {I.}~\bibnamefont {Biaggio}},\ }\href {https://doi.org/10.1103/PhysRevB.60.299} {\bibfield  {journal} {\bibinfo  {journal} {Phys. Rev. B}\ }\textbf {\bibinfo {volume} {60}},\ \bibinfo {pages} {299} (\bibinfo {year} {1999})}\BibitemShut {NoStop}%
\bibitem [{\citenamefont {Frost}(2017)}]{77}%
  \BibitemOpen
  \bibfield  {author} {\bibinfo {author} {\bibfnamefont {J.~M.}\ \bibnamefont {Frost}},\ }\href {https://doi.org/10.1103/PhysRevB.96.195202} {\bibfield  {journal} {\bibinfo  {journal} {Phys. Rev. B}\ }\textbf {\bibinfo {volume} {96}},\ \bibinfo {pages} {195202} (\bibinfo {year} {2017})}\BibitemShut {NoStop}%
\bibitem [{\citenamefont {Franchini}\ \emph {et~al.}(2021)\citenamefont {Franchini}, \citenamefont {Reticcioli}, \citenamefont {Setvin},\ and\ \citenamefont {Diebold}}]{80}%
  \BibitemOpen
  \bibfield  {author} {\bibinfo {author} {\bibfnamefont {C.}~\bibnamefont {Franchini}}, \bibinfo {author} {\bibfnamefont {M.}~\bibnamefont {Reticcioli}}, \bibinfo {author} {\bibfnamefont {M.}~\bibnamefont {Setvin}},\ and\ \bibinfo {author} {\bibfnamefont {U.}~\bibnamefont {Diebold}},\ }\href {https://doi.org/10.1038/s41578-021-00289-w} {\bibfield  {journal} {\bibinfo  {journal} {Nat. Rev. Mater.}\ }\textbf {\bibinfo {volume} {6}},\ \bibinfo {pages} {560} (\bibinfo {year} {2021})}\BibitemShut {NoStop}%
\bibitem [{\citenamefont {Feynman}(1955)}]{81}%
  \BibitemOpen
  \bibfield  {author} {\bibinfo {author} {\bibfnamefont {R.~P.}\ \bibnamefont {Feynman}},\ }\href {https://doi.org/10.1103/PhysRev.97.660} {\bibfield  {journal} {\bibinfo  {journal} {Phys. Rev.}\ }\textbf {\bibinfo {volume} {97}},\ \bibinfo {pages} {660} (\bibinfo {year} {1955})}\BibitemShut {NoStop}%
\bibitem [{\citenamefont {Bhumla}\ \emph {et~al.}(2022)\citenamefont {Bhumla}, \citenamefont {Jain}, \citenamefont {Sheoran},\ and\ \citenamefont {Bhattacharya}}]{86}%
  \BibitemOpen
  \bibfield  {author} {\bibinfo {author} {\bibfnamefont {P.}~\bibnamefont {Bhumla}}, \bibinfo {author} {\bibfnamefont {M.}~\bibnamefont {Jain}}, \bibinfo {author} {\bibfnamefont {S.}~\bibnamefont {Sheoran}},\ and\ \bibinfo {author} {\bibfnamefont {S.}~\bibnamefont {Bhattacharya}},\ }\href {https://doi.org/10.1021/acs.jpclett.2c02852} {\bibfield  {journal} {\bibinfo  {journal} {J. Phys. Chem. Lett.}\ }\textbf {\bibinfo {volume} {13}},\ \bibinfo {pages} {11655} (\bibinfo {year} {2022})},\ \bibinfo {note} {pMID: 36503226},\ \Eprint {https://arxiv.org/abs/https://doi.org/10.1021/acs.jpclett.2c02852} {https://doi.org/10.1021/acs.jpclett.2c02852} \BibitemShut {NoStop}%
\end{thebibliography}%

\end{document}